\documentclass[aps,pra, twocolumn, groupedaddress, twoside]{revtex4-2}

\usepackage{graphicx}
\usepackage{amsfonts}
\usepackage{amsmath}
\usepackage{amssymb}

\usepackage{xcolor}
\definecolor{grey}{rgb}{0.7,0.7,0.7}
\definecolor{db}{rgb}{0,0,0.5}

\usepackage{fancyhdr}
 
\begin{document}



\title{\colorbox{db}{\parbox{\linewidth}{  \centering \parbox{0.9\linewidth}{ \textbf{\centering\Large{\color{white}{ \vskip0.2em{Generic Optical Singularities and Beam-field Phenomena due to General Paraxial Beam Reflection at a Plane Dielectric Interface}\vskip0.8em}}}}}}}



\author{Anirban Debnath}
\email[]{anirban.debnath090@gmail.com}
\affiliation{School of Physics, University of Hyderabad, Hyderabad 500046, India}

\author{Nirmal K. Viswanathan}
\email[]{nirmalsp@uohyd.ac.in}
\affiliation{School of Physics, University of Hyderabad, Hyderabad 500046, India}


\date{\today}

\begin{abstract}
\noindent{\color{grey}{\rule{0.784\textwidth}{1pt}}}
\vspace{-0.8em}

A single paraxial beam reflection at a plane dielectric interface, configured appropriately, can lead to the formation of a polarization singularity in the inhomogeneously polarized output beam-field for any central angle of incidence. In this paper we derive the necessary condition to realize this effect. We explore the phase singularity characteristics associated to this polarization-singular field; and explore the dynamics of the singularities due to controlled variations of the input polarization. 
The simulation-generated exact field information lead to the exploration of the unique Goos-H\"anchen, Imbert-Fedorov and spin shifts of the optical-singular fields and the anticipation of an exact mathematical characterization of spin-orbit interaction phenomena involved therein. 
The formation of a phase singularity independent of a polarization singularity is explained subsequently.
Interrelating these seemingly unconnected beam-field phenomena and generic optical singularities can lead to a significant and fundamental understanding of the inhomogeneously polarized beam-field; and additionally, our singularity generation method can find potential application in experimental characterization of the involved dielectric media.
{\color{grey}{\rule{0.784\textwidth}{1pt}}}
\end{abstract}


\maketitle



\tableofcontents

{\color{grey}{\noindent\rule{\linewidth}{1pt}}}


\section{Introduction} \label{Sec_Intro}

The reflection and transmission of an ideal plane electromagnetic wave at a plane isotropic dielectric interface is one of the fundamental problems in electromagnetic optics \cite{BornWolf}. However, an ideal plane wave does not physically exist; instead, a real optical beam can be decomposed into constituent ideal plane waves --- each of which can be analyzed individually to understand their reflection and transmission. The composite output beams thus generated exhibit fundamentally significant beam-field phenomena --- such as Goos-H\"anchen (GH) shift, Imbert-Fedorov (IF) shift, longitudinal and transverse spin shifts --- decades of extensive studies on which are present in the literature \cite{GH, Artmann, RaJW, AntarYM, McGuirk, ChanCC, Porras, AielloArXiv, Fedorov, Schilling, Imbert, Player, FVG, Liberman, Onoda, Bliokh2006, Bliokh2007, HostenKwiat, AielloArXiv2, Aiello2008, Merano2009, Aiello2009, Qin2011, BARev, GotteLofflerDennis, XieSHELinIF}.

A different class of special characteristics of beam-fields are the phase and polarization singularities \cite{Gbur}. A phase singularity is a point in the beam-field where the phase of the field is indeterminate. This occurs when both the real and the imaginary parts of the field are zero \cite{NyeBerry1974, Bhandari97, SGV97, PA2000, SV2001, DOP09, BNRev, Gbur}.
A $C$-point polarization singularity in the beam-field is a point where the orientation of the polarization ellipse is undefined (but handedness is defined) \cite{Gbur, Nye83b, Nye83a, Hajnal87a, Hajnal87b, NH1987, DH1994, SV2001, DOP09, DennisPS02, DennisMonstar08, NKVMonstar, NKVFiber, Vpoint}. A point containing either a $\hat{\boldsymbol{\sigma}}^+$ or a  $\hat{\boldsymbol{\sigma}}^-$ spin polarization, and surrounded by other polarizations (usually arranged in the beam-field in special patterns such as lemon, star and monstar), is a $C$-singularity point.

The formation of phase singularities due to Brewster reflection has first been identified by Barczyk et al. \cite{VortexBrewster}; and subsequently, in Refs. \cite{CLEO2020} and \cite{ADNKVBrew2021} we have examined the formation and transitional dynamics of generic polarization singularities in a Brewster-reflected paraxial beam-field.
In the present paper, we show that the above Brewster-reflection effects constitute a subset of a larger class of phenomena that occur due to any general central angle of incidence. If an optical system is configured to give, e.g., a $\hat{\boldsymbol{\sigma}}^-$ spin-polarized reflected beam in the ideal case for any given angle of incidence, the wavefront-curvature of the real beam causes the appearance of the intended $\hat{\boldsymbol{\sigma}}^-$ polarization only at the beam-center, and non-circular (approximately $\hat{\boldsymbol{\sigma}}^-$) polarizations in the surrounding region. This causes the appearance of a $C$-point polarization singularity at the beam-center. 
By appropriately configuring the input polarization, the above scheme can be generalized to create a polarization singularity anywhere in the beam-field.
We derive the necessary condition to generate such a singularity in the beam-field for a general angle of incidence; and show that the above-mentioned Brewster-reflection effects \cite{VortexBrewster, CLEO2020, ADNKVBrew2021} can be seamlessly derived as a special case of the obtained condition. Additionally, we interpret the $\hat{\boldsymbol{\sigma}}^-$-polarized $C$-singularity as an attribution of an associated phase-singular $\hat{\boldsymbol{\sigma}}^+$-polarized field --- and subsequently demonstrate the dynamics of these phase and polarization singularities due to controlled variations in the optical system. The present paper thus provides a fundamental and significant understanding on the existence of phase and polarization singularities in a general reflected paraxial beam-field for any given central angle of incidence.

Even though we use a Fresnel-coefficient-based calculation only at the central plane of incidence to explain the concerned optical singularity formations, we obtain the simulated field profiles by using our generalized reflection and transmission coefficient matrix formalism \cite{ADNKVrt2020} --- via which the exact field information of the final output beam is available. As explained in Ref. \cite{ADNKVrt2020}, the various novel beam-field properties and phenomena such as GH and IF shifts, spin shifts, geometric phase characteristics \cite{P1956, Berry1984, Berry1987, Shapere, Bliokh2008, Bliokh2009, BA2010} --- and spin-orbit interaction (SOI) phenomena in general \cite{AllenOAM1992, Liberman, Berry1998, Barnett2022, Onoda, HostenKwiat, Bliokh2008, Bliokh2009, BA2010} --- are different manifestations of the same fundamental inhomogeneously polarized nature of the beam-field; and all these phenomena are exactly characterizable by using the available complete field information. By using the computationally generated exact field information, we demonstrate the GH, IF and spin shift phenomena in the context of the presently considered optical-singular fields, and explore their unique variations due to the variation of the optical system configuration. 
We then propose that the exact field information can lead to a complete mathematical characterization of the SOI characteristics of the beam-field, which are anticipated to be significant and fundamentally interesting especially in the context of the presently considered optical-singular fields. Subsequently, we explain how a phase-singular field can be achieved independent of a polarization singularity, leading to a special decomposition of an inhomogeneously polarized field as a superposition of a dominant plane-wave or near-plane-wave field and a remnant orthogonally polarized phase-singular field. Such a special decomposition is anticipated to have strong interrelation to the special beam-field phenomena.
To summarize, the special beam-field phenomena studied in the current literature and the optical-singular phenomena discussed in the present paper are all manifestations of the same fundamental polarization inhomogeneity of the beam-field; and hence all these phenomena are interrelated. We explore some of these interrelations in the present paper; and further interrelations, including a detailed mathematical characterization of SOI phenomena of the present optical-singular fields, is to be explored in the future.
In addition, our singularity formation method is anticipated to have potential application in experimental measurements of refractive indices of dielectric media.


\section{The Optical System and Field Transformation} \label{Sec_System}

\begin{figure}
\includegraphics[width = \linewidth]{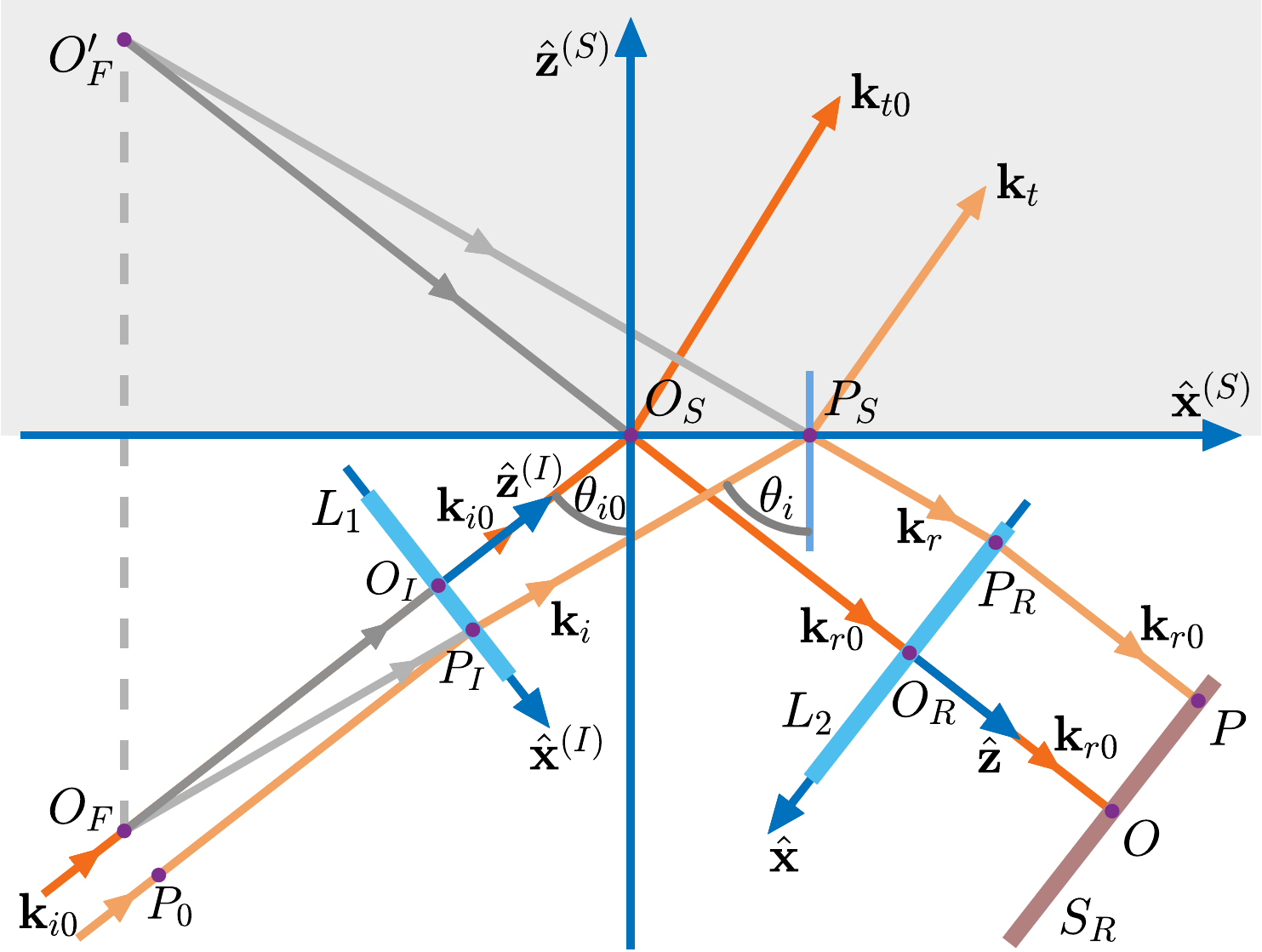}
\caption{The simulated optical system to analyze the reflection of a paraxially diverging optical beam (description in the text).}
{\color{grey}{\rule{\linewidth}{1pt}}}\label{Fig_System}
\end{figure}

We simulate an optical system [Fig. \ref{Fig_System}] based on the one we have used in Ref. \cite{ADNKVrt2020} (the same coordinate system conventions are implied). In this system, an initial collimated Gaussian beam is diverged through a lens $L_1$ (focal length $\mathcal{F}_1 < 0$). The resulting spherically diverging paraxial beam is incident at a plane isotropic dielectric interface with a central angle of incidence $\theta_{i0}$. The refractive indices of the incidence and transmission media are respectively $n_1$ and $n_2$. The reflected spherically diverging paraxial beam propagates to a lens $L_2$ (focal length $\mathcal{F}_2 > 0$) which collimates the beam. Finally, the beam profile is observed at the screen $S_R$. 
As shown in Fig. \ref{Fig_System}, $O_F$ is the focus of $L_1$ ($\mathcal{F}_1 = -O_F O_I$); and $O_F'$ is the image of $O_F$ (the beam-field never truly exists in these regions; so we validly consider $O_F$ and $O_F'$ as point-sources of the incident and reflected diverging beams respectively for the purpose of the geometrical analysis). So, the distances $O_I O_S$ and $O_S O_R$ are adjusted to get $O_F O_I + O_I O_S + O_S O_R = \mathcal{F}_2$ --- ensuring the final collimation.

We need to consider only the central plane of incidence ($y^{(I)} = y^{(S)} = y = 0$) for the present paper. In this plane, we consider an arbitrary ray path $P_0 \! \rightarrow \! P_I \! \rightarrow \! P_S \! \rightarrow \! P_R \! \rightarrow \! P$ [Fig. \ref{Fig_System}], along which a set of constituent wavefront surface elements are considered to propagate. If the $x^{(I)}$ coordinate of $P_I$ is $x_I$, then the $x$ coordinate of $P_R$ is obtained by the system geometry as $x_R = -\alpha x_I$, where $\alpha = \mathcal{F}_2/|\mathcal{F}_1|$. The same geometry also shows that, as an associated spherical surface element propagates from $P_I$ to $P_R$, its area expands by a factor $\alpha^2$ --- due to inverse square law.

We now consider an input surface-element field at $P_0$ as (suppressing the $\mathbf{k}\cdot\mathbf{r} - \omega t$ phase term)
\begin{subequations}
\begin{eqnarray}
& \boldsymbol{\mathcal{E}}_0^{(I)} = \mathcal{E}_{0x}^{(I)} \, \hat{\mathbf{x}}^{(I)} + e^{i\Phi_E} \mathcal{E}_{0y}^{(I)} \, \hat{\mathbf{y}}^{(I)}; & \label{E0I} \\
& \mathcal{E}_{0x}^{(I)} = \mathcal{E}_{00} \, G_I \cos\theta_E; \hspace{1em} \mathcal{E}_{0y}^{(I)} = \mathcal{E}_{00} \, G_I \sin\theta_E; & \label{E0x,E0y}
\end{eqnarray}
\end{subequations}
where, $\mathcal{E}_{00}$ represents the central field magnitude; $G_I = e^{-x_I^{2}/w_0^2}$ represents the Gaussian distribution along the $x^{(I)}$ axis with a half-width $w_0$; and $(\theta_E, \Phi_E)$ represent the angle and relative-phase parameters determining the field polarization. 
As this element field propagates to $P$, it is (1) unaltered along $P_0 \! \rightarrow \! P_I$ and $P_R \! \rightarrow \! P$; (2) modified by the lenses $L_1$ and $L_2$ at $P_I$ and $P_R$ respectively; (3) reduced in amplitude by a factor $g = 1/\alpha$ due to the surface-element area expansion along $P_I \! \rightarrow \! P_S \! \rightarrow \! P_R$; and (4) modified by Fresnel reflection coefficients at $P_S$. We have analytically verified that, at the central plane of incidence, the modifications due to $L_1$ and $L_2$ exactly compensate for each other; and hence, their exact analysis is not required here.

At the central plane of incidence, the $\hat{\mathbf{x}}^{(I)}$ and $\hat{\mathbf{y}}^{(I)}$ components of $\boldsymbol{\mathcal{E}}_0^{(I)}$ [Eq. (\ref{E0I})] are respectively the transverse magnetic (TM) and transverse electric (TE) components. Hence, at $P_S$ these components acquire Fresnel TM and TE reflection coefficients $r_{TM}(\theta_i)$ and $r_{TE}(\theta_i)$ respectively \cite{BornWolf},
where, $\theta_i$ is the angle of incidence at $P_S$. The final output field at $P$ is thus obtained as
\begin{subequations}
\begin{eqnarray}
& \boldsymbol{\mathcal{E}} = \mathcal{E}_{x} \, \hat{\mathbf{x}} + e^{i\Phi_E} \mathcal{E}_{y} \, \hat{\mathbf{y}}; & \label{ER} \\
& \mathcal{E}_{x} = \mathcal{E}_{R} \, r_{TM} (\theta_{i}) \cos\theta_E; \hspace{1em} \mathcal{E}_{y} = \mathcal{E}_{R} \, r_{TE} (\theta_{i}) \sin\theta_E; \hspace{0.5em} & \label{ExR_EyR} \\
& \mathcal{E}_R = g \, \mathcal{E}_{00} \, G_R; \hspace{0.8em} G_I \equiv G_R = e^{-x_R^{2}/w_R^2}; \hspace{0.5em} w_R = \alpha \, w_0. \hspace{1.4em} & \label{GR_def}
\end{eqnarray}
\end{subequations}
Since the ray path $P_0 \! \rightarrow \! P$ is arbitrary,
the field $\boldsymbol{\mathcal{E}}$ [Eq. (\ref{ER})] truly represents the final output electric field as a function of $x \equiv x_R$ at the linear section of the screen $S_R$ at the central plane of incidence.


\section{Polarization Singularity Formation} \label{Sec_PolSing}

\subsection{Generalized Condition}


As understood from Eq. (\ref{ExR_EyR}), the Gaussian function $G_R$ is modified by $r_{TM} (\theta_{i})$ and $r_{TE} (\theta_{i})$ respectively to generate different field functions $\mathcal{E}_x(x)$ and $\mathcal{E}_y(x)$.
However, for a given specific point $x = x_S$ in the beam-field, it is possible to find a specific value $\theta_E = \theta_{ES}$ that satisfies
\begin{equation}
|\tan\theta_{ES}| = |r_{TM}(\theta_{i})/r_{TE}(\theta_{i})|_{x = x_S}, \label{mod_tan_basic}
\end{equation}
resulting in $|\mathcal{E}_x| = |\mathcal{E}_y|$ only at $x = x_S$. 
Under this condition, either a $\hat{\boldsymbol{\sigma}}^+$ or a  $\hat{\boldsymbol{\sigma}}^-$ spin polarization can be generated at $x = x_S$ by choosing $\Phi_E = \pm \pi/2$ in Eq. (\ref{ER}). Equation (\ref{ExR_EyR}) ensures that another $\hat{\boldsymbol{\sigma}}^\pm$ polarization do not appear in the immediate vicinity of $x = x_S$, since $|\mathcal{E}_x| = |\mathcal{E}_y|$ is not satisfied at other points in the immediate vicinity. 

While the condition of Eq. (\ref{mod_tan_basic}) is written based on the field functions at the $x$ axis only [Eq. (\ref{ExR_EyR})], we have computationally verified based on the formalism of Ref. \cite{ADNKVrt2020} that no other immediately neighbouring point at the screen $S_R$ ($xy$ plane) in general contains a $\hat{\boldsymbol{\sigma}}^\pm$ polarization under the above condition. An isolated $\hat{\boldsymbol{\sigma}}^\pm$ polarization, i.e. a $C$-point singularity, is thus obtained at the specific point $x = x_S$.

Thus, summarizing the central result of the present paper, a polarization singularity is obtained at a point $P(x_S,0)$ at the screen $S_R$, for any central angle of incidence $\theta_{i0}$, if the initial input polarization parameter values 
$(\theta_E, \Phi_E)$ are chosen as $(\theta_{ES}, \Phi_{ES})$, where
\begin{equation}
\tan\theta_{ES} = \pm \left|\dfrac{r_{TM}(\theta_{i})}{r_{TE}(\theta_{i})}\right|_{x = x_S}; \hspace{2em} \Phi_{ES} = \pm \dfrac{\pi}{2}. \label{Condition}
\end{equation}
This result reveals the significant fact that, when configured appropriately, even one simple reflection of a paraxial Gaussian beam at a plane dielectric interface, for any angle of incidence, can generate a polarization singularity in the reflected beam-field.

\subsection{Functional Variation of $\theta_{ES}$}

\begin{figure}
\includegraphics[width = 0.9\linewidth]{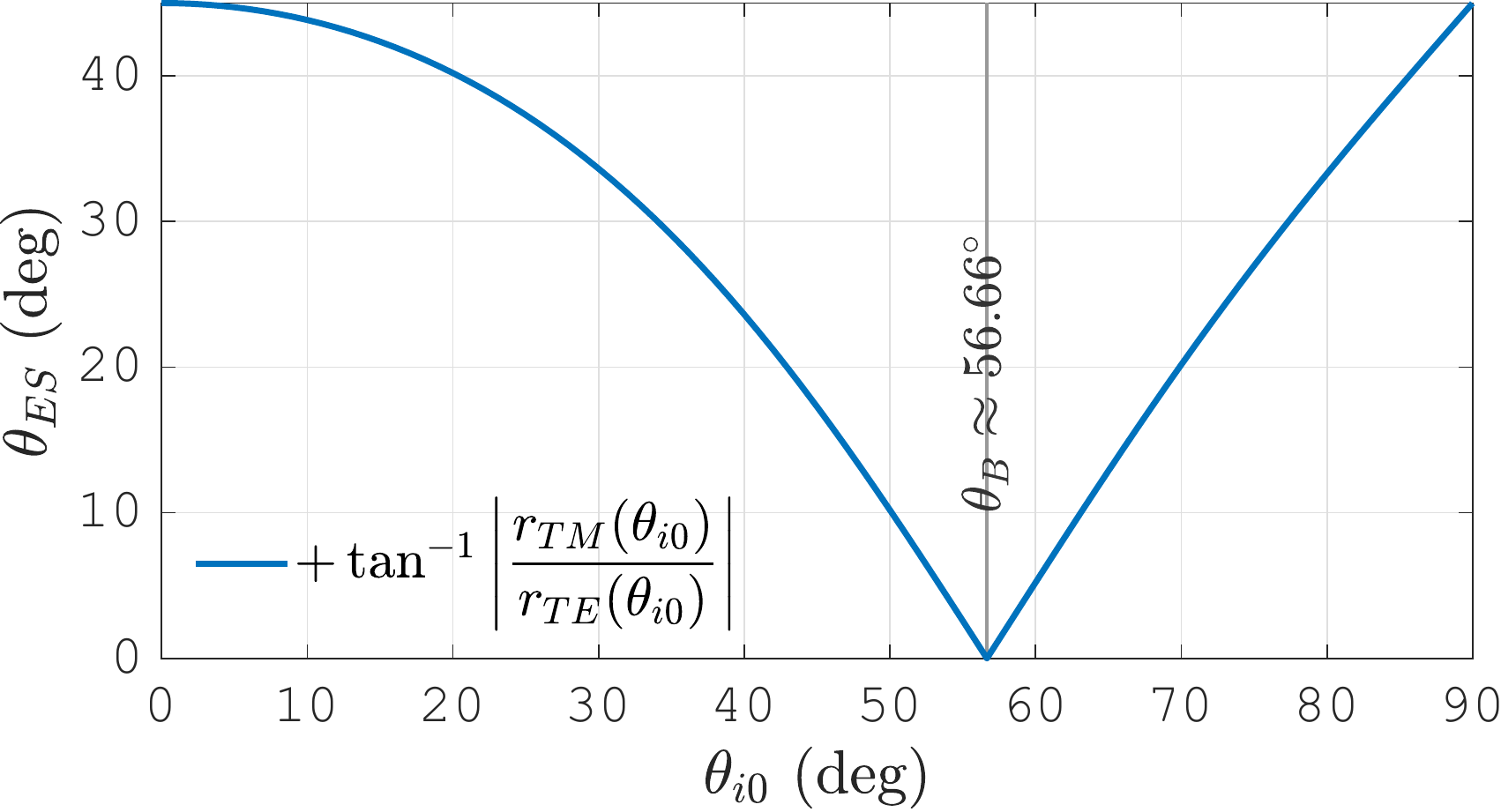}
\caption{Variation of $\theta_{ES}$ as a function of $\theta_{i0}$ [Eq. (\ref{tthES0})], considering the singularity generation at the reflected beam center. The plot shows only the positive solution $\theta_{ES} = +\tan^{-1}| r_{TM}(\theta_{i0}) / r_{TE}(\theta_{i0}) |$ for convenience; while both $\pm\tan^{-1}| r_{TM}(\theta_{i0}) / r_{TE}(\theta_{i0}) |$ are valid solutions.}
{\color{grey}{\rule{\linewidth}{1pt}}}
\label{Fig_thEplot}
\end{figure}

For the purpose of simulations and experimental results in the present paper, it is convenient to show the formation of a polarization singularity at the central-ray point $O$ ($x_S = 0$) at the screen $S_R$ [Fig. \ref{Fig_System}]. For $x_S = 0$, Eq. (\ref{Condition}) gives
\begin{subequations}
\begin{eqnarray}
& \tan\theta_{ES} = \pm\left|\dfrac{r_{TM}(\theta_{i0})}{r_{TE}(\theta_{i0})}\right| = 
\pm\left|\dfrac{\cos(\theta_{i0} + \theta_{t0})}{\cos(\theta_{i0} - \theta_{t0})}\right|; & \label{tthES0} \\
& \theta_{t0} = \sin^{-1} \left[(n_1/n_2) \sin\theta_{i0}\right], \hspace{1em} (\mbox{Snell's law}); & 
\end{eqnarray}
\end{subequations}
obtained using the Fresnel coefficient expressions \cite{BornWolf} ($c_i = \cos\theta_{i0}$, $c_t = \cos\theta_{t0}$)
\begin{subequations}
\begin{eqnarray}
& r_{TM}(\theta_{i0}) = (n_2 c_i - n_1 c_t)/(n_2 c_i + n_1 c_t); & \\
& r_{TE}(\theta_{i0}) = (n_1 c_i - n_2 c_t)/(n_1 c_i + n_2 c_t). & 
\end{eqnarray}
\end{subequations}

The variation of $\theta_{ES}$ as a function of $\theta_{i0}$ [Eq. (\ref{tthES0})], for $n_1 = 1$ and $n_2 = 1.52$, is shown in Fig. \ref{Fig_thEplot}. Some of the notable characteristics of this variation are as follows:

\begin{enumerate}

\item For $\theta_{i0} = 0^\circ$, we have $|r_{TM} (0^\circ)| = |r_{TE} (0^\circ)| = (n_2 - n_1)/(n_2 + n_1)$. So, to satisfy the condition of Eq. (\ref{tthES0}), we have $\theta_{ES} = \pm 45^\circ$.

\item For $\theta_{i0} \rightarrow 90^\circ$, we have $r_{TM} (\theta_{i0}) \approx r_{TE} (\theta_{i0}) \rightarrow -1$. Hence we have $\theta_{ES} \rightarrow \pm 45^\circ$ to satisfy the condition of Eq. (\ref{tthES0}). 

\item For Brewster angle incidence $\theta_{i0} = \theta_B = \tan^{-1}(n_2/n_1)$, we have $r_{TM} (\theta_{B}) = 0$. Thus, to satisfy Eq. (\ref{tthES0}), we must have $\theta_{ES} = 0^\circ$. This is the only special case where, according to Eq. (\ref{ExR_EyR}), both $\mathcal{E}_x$ and $\mathcal{E}_y$ must become zero at $O$ to create a singularity. It is thus a higher-order node-singularity \cite{Gbur}, but not a simple $C$-point singularity. The formation of this node-singularity, however, can be explained in terms of a merger of two $C$-point singularities, as we have explained in our earlier work of Ref. \cite{ADNKVBrew2021}.

\end{enumerate}

Figure \ref{Fig_thEplot} clearly shows that
the formation of a polarization singularity due to Brewster-reflection is seamlessly identified by considering $\theta_{i0} = \theta_B$ in the general condition of Eq. (\ref{tthES0}) --- which is a remarkable result, since this generalization has not been anticipated in the previous works \cite{VortexBrewster, CLEO2020, ADNKVBrew2021}. Our entire work of Ref. \cite{ADNKVBrew2021} can thus be considered as a special-case analysis of the present work on generalized singularity generation due to paraxial beam reflection.


\section{Simulations and Experiments} \label{Sec_SimExp}

\subsection{Experimental Setup}

\begin{figure}
\includegraphics[width = \linewidth]{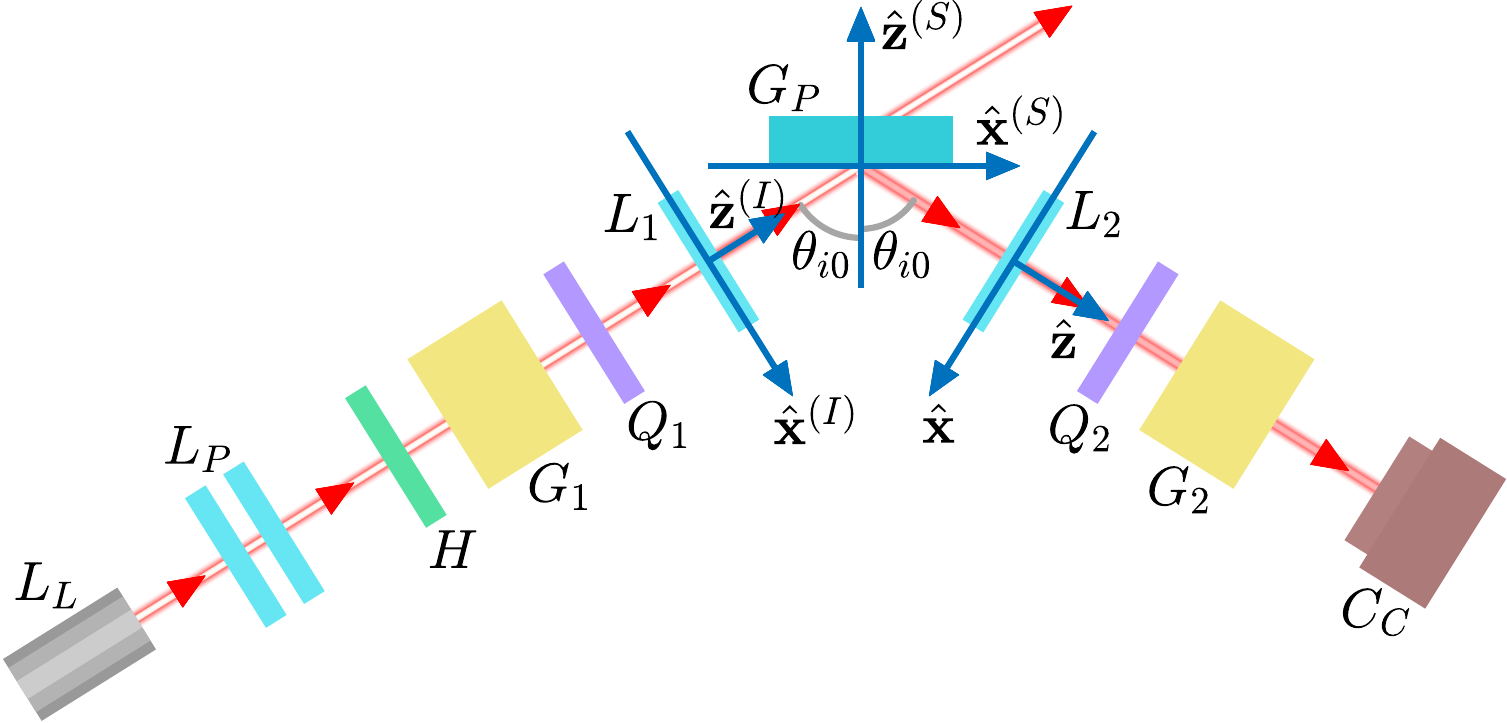}
\caption{The experimental setup, comprising of a He-Ne laser $L_L$, a collimating lens-pair $L_P$, a half wave plate $H$, Glan-Thompson polarizers $G_1$ and $G_2$, quarter wave plates $Q_1$ and $Q_2$, diverging and collimating lenses $L_1$ and $L_2$, a glass plate $G_P$ whose surface is used as the dielectric interface, and a CCD camera $C_C$ which represents the screen of observation $S_R$ [Fig. \ref{Fig_System}] (adapted from our previous work of Ref. \cite{ADNKVBrew2021}).}
{\color{grey}{\rule{\linewidth}{1pt}}}
\label{Fig_ExpSetup}
\end{figure}



For the purpose of the present paper, we experimentally recreate the simulated system of Fig. \ref{Fig_System}, as shown and briefly captioned in Fig. \ref{Fig_ExpSetup} (adapted from our previous work of Ref. \cite{ADNKVBrew2021}). 
In addition to the components of Fig. \ref{Fig_System}, the setup also includes a quarter wave plate (QWP) and a Glan-Thompson polarizer (GTP) after the lens $L_2$ in order to make Stokes measurements \cite{Goldstein} on the final output beam. The various parameters considered for the simulation and the experiment are: refractive indices $n_1 = 1$, $n_2 = 1.52$; power and free-space wavelength of the laser, $P_w = 1$ mW, $\lambda = 632.8$ nm; half-width of the input beam, $w_0 = 0.6$ mm; focal lengths of the lenses, $\mathcal{F}_1 = -5$ cm, $\mathcal{F}_2 = 12.5$ cm; propagation distances $O_I O_S = 5$ cm, $O_S O_R = 2.5$ cm.



\subsection{Simulated and Experimental Examples}

\begin{figure*}
\includegraphics[width = 0.78\linewidth]{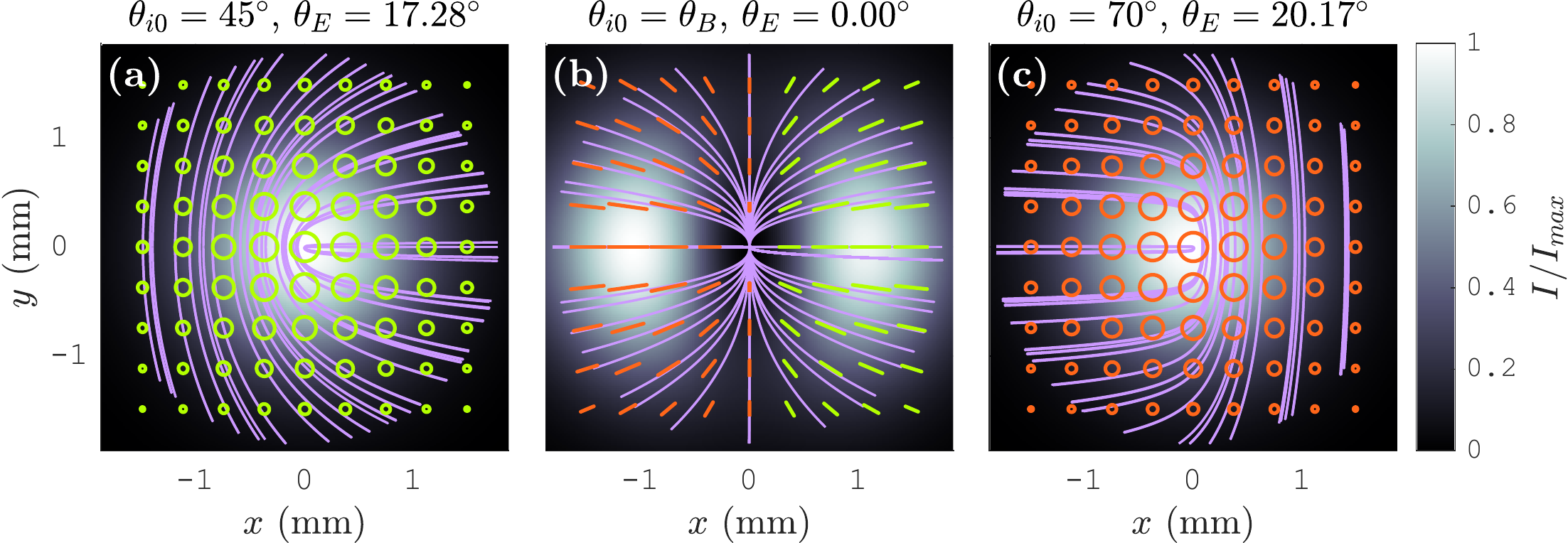}\\
\caption{Simulated polarization-singular field profiles for \textbf{(a)} $\theta_{i0} = 45^\circ$ and $\theta_E \approx 17.28^\circ$ (obtained $I_{max} \approx 4.83$ W/m$^2$); \textbf{(b)} $\theta_{i0} = \theta_B \approx 56.66^\circ$ and $\theta_E = 0^\circ$ (obtained $I_{max} \approx 2.93$ mW/m$^2$); and \textbf{(c)} $\theta_{i0} = 70^\circ$ and $\theta_E \approx 20.17^\circ$ (obtained $I_{max} \approx 20.73$ W/m$^2$). 
For each profile, we have chosen $\Phi_E = +\pi/2$.
\textbf{Color code:} Light green and dark orange ellipses respectively represent left-elliptical (i.e. $\hat{\boldsymbol{\sigma}}^-$-dominant) and right-elliptical (i.e. $\hat{\boldsymbol{\sigma}}^+$-dominant) polarizations.}
{\color{grey}{\rule{\linewidth}{1pt}}}
\label{Fig_EProfilesTh}
\end{figure*}

\begin{figure}
\includegraphics[width = \linewidth]{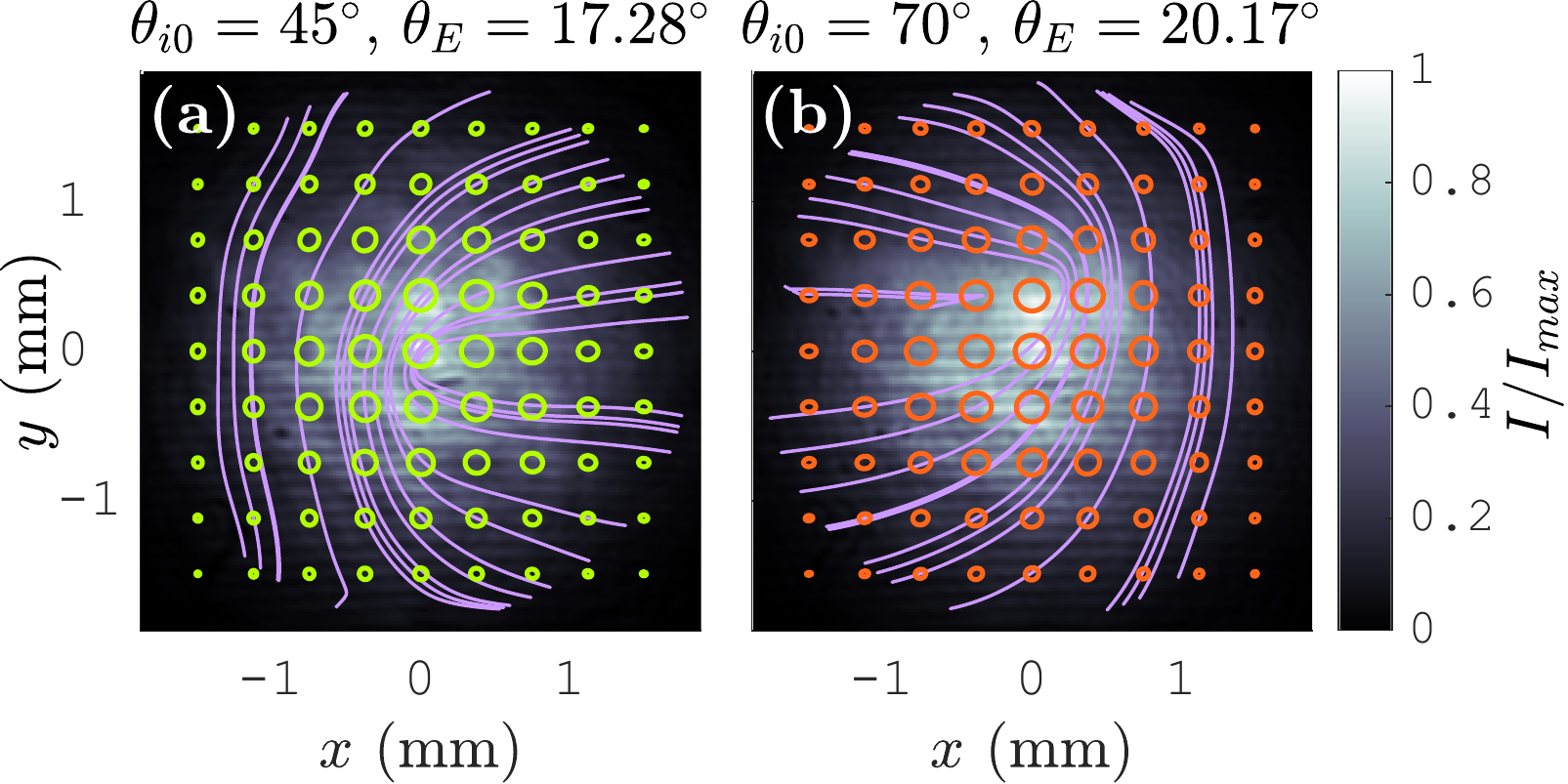}\\
\caption{Experimentally obtained field profiles corresponding to the simulated profiles of Figs. \ref{Fig_EProfilesTh}(a) and \ref{Fig_EProfilesTh}(c); showing a reasonable match between the simulations and experiments (experimental verification of the profile of Fig. \ref{Fig_EProfilesTh}(b) can be found in Ref. \cite{ADNKVBrew2021}).}
{\color{grey}{\rule{\linewidth}{1pt}}}
\label{Fig_EProfilesExp}
\end{figure}


We have computationally generated field profiles for various $\theta_{i0}$ values with the condition of Eq. (\ref{tthES0}) applied, including the $\theta_{i0} = \theta_B$ case, and have obtained the expected polarization singularities. 
Simulated examples of polarization-singular field profiles for $\theta_{i0} = 45^\circ$, $56.66^\circ$ ($\theta_B$) and $70^\circ$ are shown in Fig. \ref{Fig_EProfilesTh}.  
Corresponding to these $\theta_{i0}$ values, the $\theta_{ES}$ values are obtained from Eq. (\ref{tthES0}) as $\theta_{ES} \approx \pm 17.28^\circ$, $0^\circ$ and $\pm 20.17^\circ$. The profiles of Fig. \ref{Fig_EProfilesTh} 
are obtained by considering only the positive parameter values $\theta_E = +|\theta_{ES}|$ and $\Phi_E = +\pi/2$. The profiles for the negative parameter values are straightforward to obtain, and are not shown here.

The profile for the $\theta_{i0} = \theta_B$ case [Fig. \ref{Fig_EProfilesTh}(b)] matches well with the result of our earlier work of Ref. \citep{ADNKVBrew2021}, showing the formation of a node singularity. The profiles of Figs. \ref{Fig_EProfilesTh}(a) and \ref{Fig_EProfilesTh}(c) show examples of a $\theta_{i0} < \theta_B$ case and a $\theta_{i0} > \theta_B$ case, which contain a central $\hat{\boldsymbol{\sigma}}^-$ polarization and a central $\hat{\boldsymbol{\sigma}}^+$ polarization respectively, associated to opposite lemon patterns. We have observed in the simulation that, by transforming either $\theta_{E} \rightarrow -\theta_E$ or $\Phi_{E} \rightarrow -\Phi_E$, the handedness of the central spin polarization can be flipped; whereas, changing both the signs keeps the handedness unchanged.


Figure \ref{Fig_EProfilesExp} shows the experimentally-generated field profiles corresponding to the simulated profiles of Figs. \ref{Fig_EProfilesTh}(a) and \ref{Fig_EProfilesTh}(c) respectively (the $\theta_{i0} < \theta_B$ and $\theta_{i0} > \theta_B$ cases; the experimental verification for $\theta_{i0} = \theta_B$ [Fig. \ref{Fig_EProfilesTh}(b)] can be found in Ref. \cite{ADNKVBrew2021}). We have obtained these experimental field profiles by analyzing the results of Stokes measurements \cite{Goldstein} performed on the output beam [Fig. \ref{Fig_ExpSetup}]. The experimental profiles match well with the corresponding simulated profiles --- thus verifying the formation of the expected polarization singularities.

In the light of the above analysis, it is evident that the formation of an isolated $C$-point singularity in Figs. 4(c) and 6(c) of Ref. \cite{ADNKVBrew2021} is simply an example of an off-Brewster incidence case of the present generalized formalism --- which is another remarkable result here.

\subsection{Phase Singularity of a Spin-Component Field} \label{Subsec_PhaseVortex}


\begin{figure}
\includegraphics[width = \linewidth]{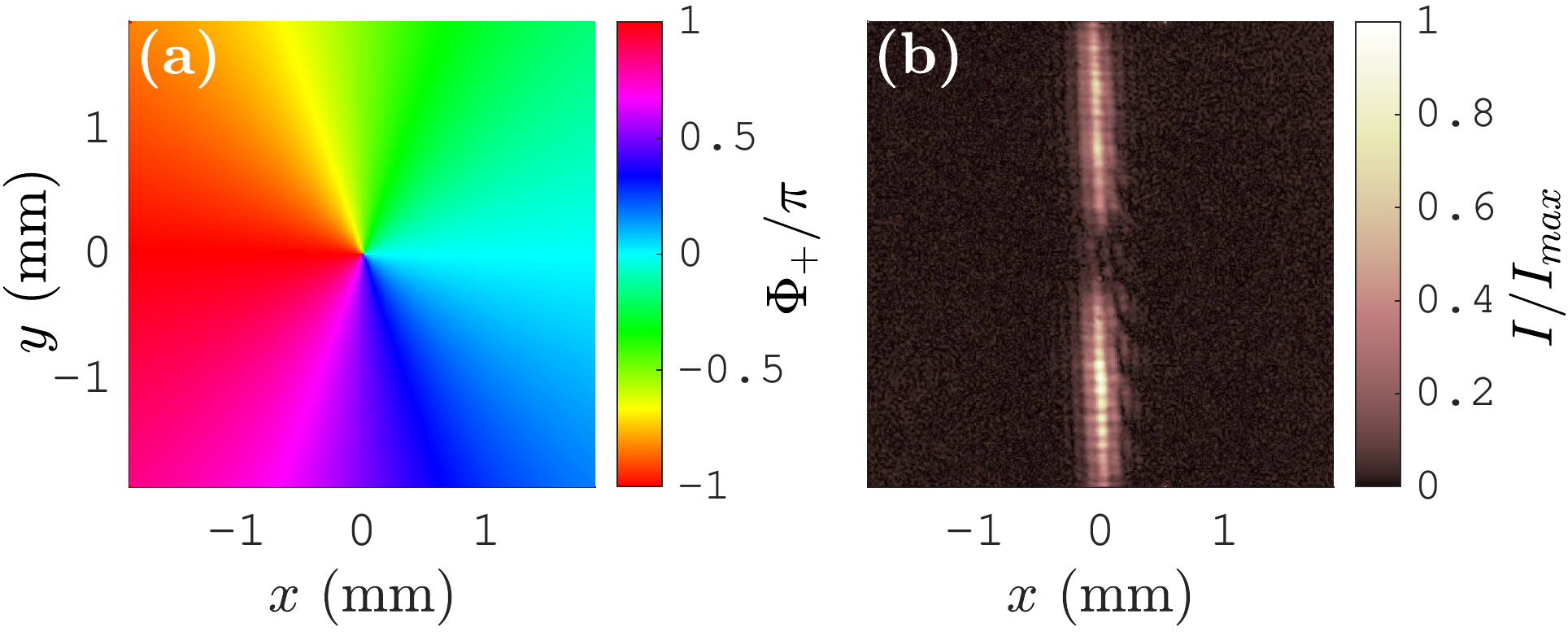} 
\caption{\textbf{(a)} Phase profile of the $\hat{\boldsymbol{\sigma}}^+$-polarized component field corresponding to the polarization-singular total beam-field of Fig. \ref{Fig_EProfilesTh}(a), exhibiting an $l \! = \! -1$ phase vortex. \textbf{(b)} Single slit diffraction pattern of the $\hat{\boldsymbol{\sigma}}^+$-converted-to-$\hat{\mathbf{d}}^+$ controlled output beam, showing the characteristic fringe-dislocation due to an $l \! = \! -1$ phase vortex.}
{\color{grey}{\rule{\linewidth}{1pt}}}\label{Fig_IPhpm}
\end{figure}

Since a $C$-singularity point is a point of an isolated circular polarization surrounded by non-circular polarizations, it is implied that the contribution of the orthogonal circular polarization is precisely zero at that point and non-zero at the surrounding points. This indicates that, at the $C$-singularity point, the orthogonal circular polarization contains a phase singularity (vortex). This is exactly the way how a phase singularity and a polarization singularity are interrelated --- a circularly polarized field with a constant (or almost constant) phase profile superposes with the orthogonal-circularly polarized field with a phase singularity to generate a $C$-point polarization singularity \cite{Gbur, NKVMonstar, ADNKVBrew2021}.

To examine the phase singularity characteristics in the present work, we consider the beam-field profile for $(\theta_{i0}, \theta_{E}, \Phi_{E}) = (45^\circ, 17.28^\circ, \pi/2)$ [Figs. \ref{Fig_EProfilesTh}(a), \ref{Fig_EProfilesExp}(a)] that contains a central $C$-singularity with a $\hat{\boldsymbol{\sigma}}^-$ spin polarization. This implies that the $\hat{\boldsymbol{\sigma}}^+$-polarized component field must contain a phase singularity at the beam-center in this case.


To isolate the $\hat{\boldsymbol{\sigma}}^+$ contribution to the total field, we first propagate the output beam through a QWP,
oriented to introduce an additional $+\pi/2$ phase to the $\hat{\mathbf{x}}$ component.
This transforms the $\hat{\boldsymbol{\sigma}}^\pm$ polarizations to $\hat{\mathbf{d}}^\pm = (\hat{\mathbf{x}} \pm \hat{\mathbf{y}})/\sqrt{2}$ polarizations. Then, by passing the QWP-output beam through a GTP, with its transmission axis oriented along $\hat{\mathbf{d}}^+$, a controlled $\hat{\mathbf{d}}^+$-polarized output beam is obtained --- whose intensity and phase profiles are equivalent to those of the original $\hat{\boldsymbol{\sigma}}^+$ component field. The properties of this controlled output beam are then easily examined to reveal the properties of the original $\hat{\boldsymbol{\sigma}}^+$-polarized field.

The simulated phase profile of the $\hat{\boldsymbol{\sigma}}^+$ component field is shown in Fig. \ref{Fig_IPhpm}(a), which reveals the existence of an $l \! = \! -1$ phase vortex at the beam-center. We experimentally verify this phase characteristic by obtaining a single-slit diffraction pattern. We pass the controlled output beam through a single vertical slit of width 0.4 mm, placed along $x = 0$. The obtained far-field diffraction pattern [Fig. \ref{Fig_IPhpm}(b)] is the characteristic single-slit diffraction pattern of an $l \! = \! -1$ phase vortex \cite{SingleSlit}. This observation verifies the existence of the phase vortex in the controlled output beam.


Thus, by using the methods of the present subsection, the phase vortex characteristics of the polarization-singular output beam-field are analyzed and experimentally verified. This vortex nature, and thus an orbital angular momentum (OAM), clearly manifests itself due to the inhomogeneous nature of the reflection process, even when the initial input beam contains only spin angular momentum. This indicates towards some amount of spin-to-orbital angular momentum conversion --- a manifestation of SOI --- happening in the system due to the reflection process.

\subsection{Variation of $\theta_{E}$ for a Fixed $\theta_{i0}$} \label{Subsec_thEVariation}

\begin{figure*}
\includegraphics[width = 0.68\linewidth]{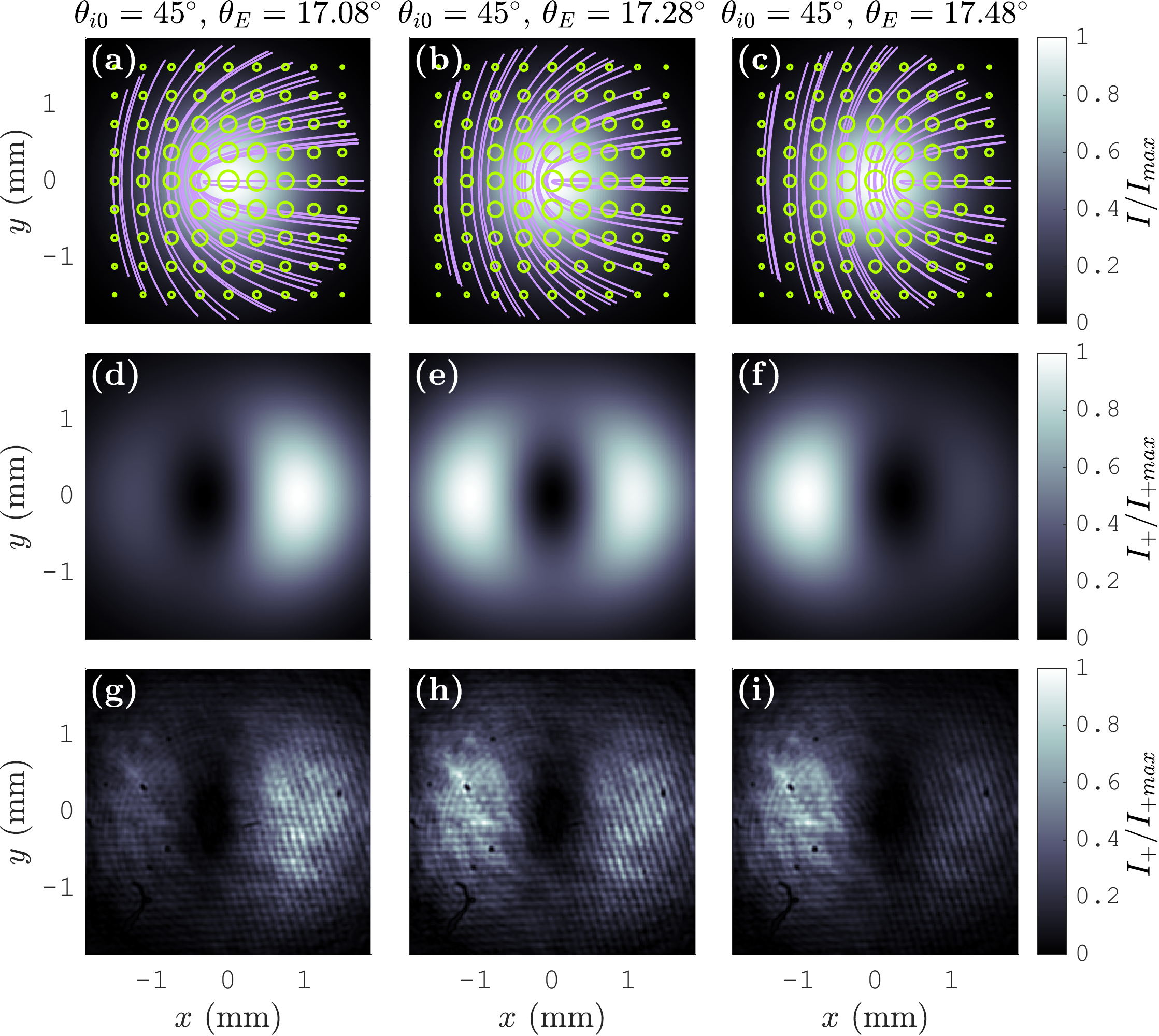}
\caption{\textbf{(a, b, c)} Simulated polarization-singular total field profiles for $\theta_{i0} = 45^\circ$ and $\theta_E \approx 17.08^\circ$, $17.28^\circ$, $17.48^\circ$ --- showing the singularity positions in the $x < 0$ region, at $x = 0$, and in the $x > 0$ region respectively. The obtained $I_{max}$ values are respectively 4.78 W/m$^2$, 4.83 W/m$^2$ and 4.88 W/m$^2$.
\textbf{(d, e, f)} Simulated intensity ($I_+$) profiles of the phase-singular $\hat{\boldsymbol{\sigma}}^+$ component field corresponding to (a), (b) and (c). The obtained $I_{+max}$ values are respectively 1.25 mW/m$^2$, 0.74 mW/m$^2$ and 1.29 mW/m$^2$. 
\textbf{(g, h, i)} Corresponding experimentally observed intensity profiles of the $\hat{\boldsymbol{\sigma}}^+$-converted-to-$\hat{\mathbf{d}}^+$ controlled output field. 
}
{\color{grey}{\rule{\linewidth}{1pt}}}\label{Fig_Ipm}
\end{figure*}

Until here, for a fixed $\theta_{i0}$ value, we have used $\theta_{E} = \theta_{ES}$ determined by Eq. (\ref{tthES0}) to obtain the intended singularity at $x = 0$. In this subsection, we vary $\theta_{E}$ around this central $\theta_{ES}$ value for the fixed $\theta_{i0}$ to observe the shift of the singularity from the beam-center, following the general condition of Eq. (\ref{Condition}).


We choose the fixed value $\theta_{i0} = 45^\circ$ as in Figs. \ref{Fig_EProfilesTh}(a) and \ref{Fig_IPhpm}. We then choose displaced $\theta_{E}$ values $17.08^\circ$ and $17.48^\circ$, which are $\pm 0.2^\circ$ shifted from the previously used value $17.28^\circ$ that produces the central singularity [Figs. \ref{Fig_EProfilesTh}(a) and \ref{Fig_IPhpm}(a)]. 
The simulated polarization-singular total beam-fields for $\theta_{E} \approx 17.08^\circ$, $17.28^\circ$ and $17.48^\circ$ are shown in Figs. \ref{Fig_Ipm}(a), \ref{Fig_Ipm}(b) and \ref{Fig_Ipm}(c) respectively, which show the singularity positions in the $x < 0$ region, at $x = 0$, and in the $x > 0$ region.
The phase-singular $\hat{\boldsymbol{\sigma}}^+$ component field also undergoes corresponding transformations --- as understood from its case-wise simulated intensity profiles of Figs. \ref{Fig_Ipm}(d), \ref{Fig_Ipm}(e) and \ref{Fig_Ipm}(f) --- because the polarization singularity of the total field and the phase singularity of the $\hat{\boldsymbol{\sigma}}^+$ field appear at the same point.
The case-wise experimentally observed $\hat{\boldsymbol{\sigma}}^+$-converted-to-$\hat{\mathbf{d}}^+$ controlled output intensity profiles are shown in Figs. \ref{Fig_Ipm}(g), \ref{Fig_Ipm}(h) and \ref{Fig_Ipm}(i) respectively, which match well with the corresponding simulated $\hat{\boldsymbol{\sigma}}^+$ intensity profiles of Figs. \ref{Fig_Ipm}(d), \ref{Fig_Ipm}(e) and \ref{Fig_Ipm}(f). 
Thus, the condition of Eq. (\ref{Condition}) is verified from a general perspective in terms of central as well as off-central singularity formations.


One can easily visualize that, as the $\theta_{E}$ value is gradually taken from $17.08^\circ$ to $17.48^\circ$ through $17.28^\circ$, the singularity position $x_S$ moves from the $x < 0$ to the $x > 0$ region through $x = 0$. Correspondingly, the Fig. \ref{Fig_Ipm}(a) (or (d), (g)) profile gradually transforms to the Fig. \ref{Fig_Ipm}(c) (or (f), (i)) profile through the intermediate profile of Fig. \ref{Fig_Ipm}(b) (or (e), (h)). This phenomenon reveals a significant transitional dynamics of the polarization and phase singularities that are demonstrated in the present paper. 




\section{Beam-shifts, Spin-shifts and a Generalized Perspective} \label{Sec_ShiftGeneral}

The central content of the present paper, as demonstrated until here, is the identification and control of generic optical singularity characteristics of a reflected paraxial beam-field; and we have used a Fresnel-coefficient-based calculation only at the central plane of incidence to identify these singularities. However, we emphasize that these singularity characteristics are a subset of a much larger set of phenomena exhibited by the concerned inhomogeneously polarized reflected beam-field. Following our work of Ref. \cite{ADNKVrt2020}, we have obtained the simulated fields 
via the generalized reflection and transmission coefficient matrix formalism, due to which the complete and exact vectorial information on the reflected beam-field is available. As discussed in Ref. \cite{ADNKVrt2020}, the availability of the complete field information serves as the foundation for determining all special beam-field properties such as GH and IF shifts, longitudinal and transverse spin shifts including the spin-Hall effect of light  (SHEL) \cite{Onoda, HostenKwiat}, and the underlying geometric phase properties of the field. 
From this perspective, we assert that all these special beam-field phenomena and the optical singularity generation discussed in the present paper are interrelated, since they are manifestations of different properties of the same inhomogeneously polarized field. In this section we demonstrate the GH, IF and spin shift phenomena of the reflected field under the considered context of singularity formation in order to show the interdependence of these seemingly unrelated effects.
And then we identify possible ways to achieve further generalization of our work from the perspective of availability of the complete field information.

\subsection{GH and IF Shifts}

\begin{figure}
\includegraphics[width = 0.9\linewidth]{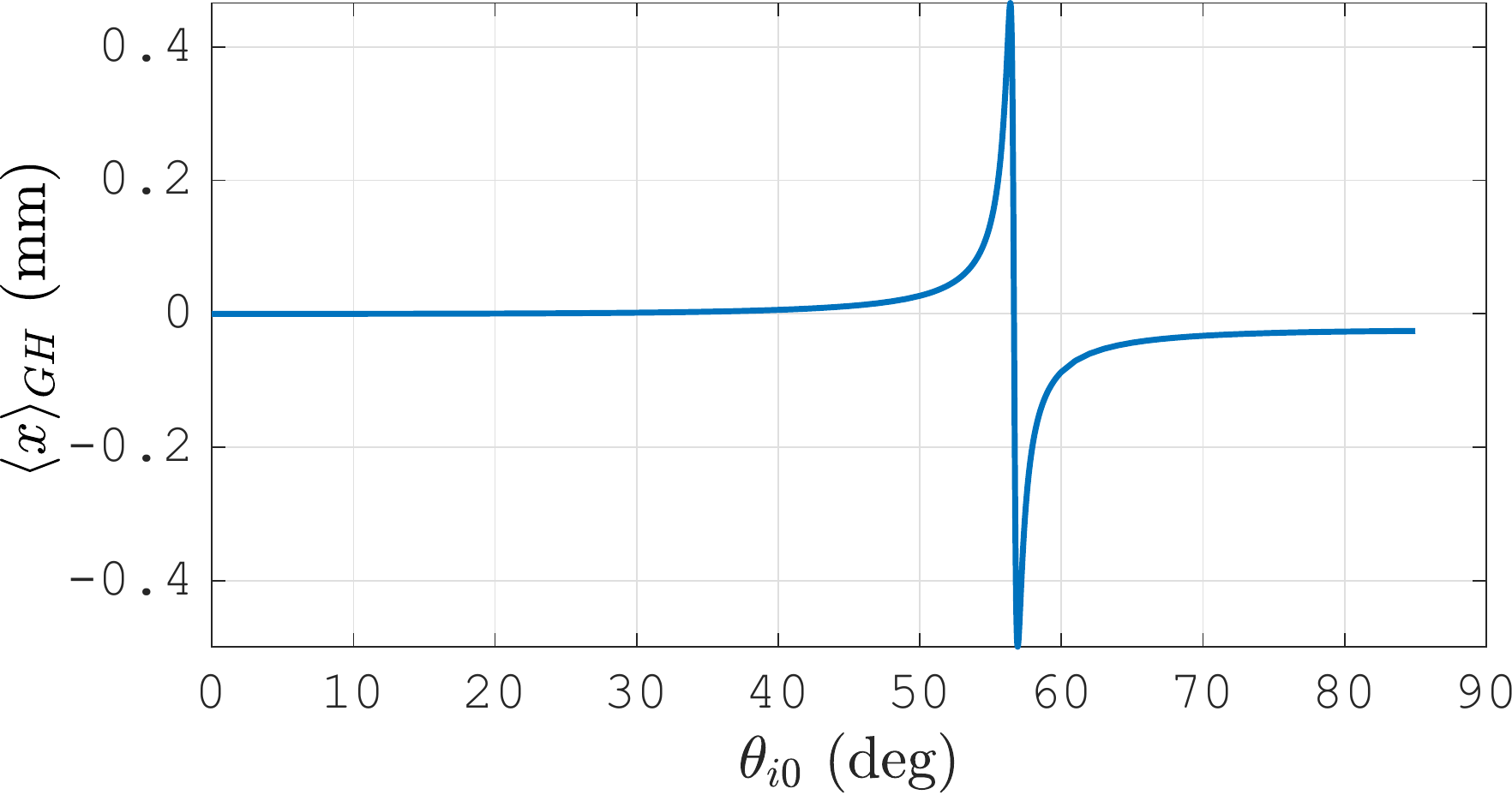}
\caption{Variation of the GH shift $\langle x\rangle_{GH}$ of the total output beam-field as a function of $\theta_{i0}$, considering the associated $\theta_{E}$ variation as $\theta_E = \theta_{ES}$ [Eq. (\ref{tthES0}), Fig. \ref{Fig_thEplot}] for singularity generation at the beam-center. The plot shows large GH shifts (comparable to the half-beam-width $w_R$) for near-Brewster incidence. A zero centroid shift is observed not exactly for the Brewster angle, but for a nearby angle $\theta_{i0} \approx 56.65^\circ$ --- due to the slight asymmetry of the Brewster-reflected intensity profile [Fig. \ref{Fig_EProfilesTh}(b)] with respect to the $y$ axis.}
{\color{grey}{\rule{\linewidth}{1pt}}}\label{Fig_thi0xGH}
\end{figure}

The GH shift of the beam is the longitudinal (i.e. along the $x$ axis) shift of the beam-centroid from the central-ray point $O$ [Fig. \ref{Fig_System}]. As discussed in Ref. \cite{ADNKVrt2020}, the reflected fields corresponding to the individual $\mathcal{E}_{0x}^{(I)} \, \hat{\mathbf{x}}^{(I)}$ and $\mathcal{E}_{0y}^{(I)} \, \hat{\mathbf{y}}^{(I)}$ input fields [Eq. (\ref{E0I})] experience their own GH shifts which vary with $\theta_{i0}$ --- and these individual GH shifts contribute to the GH shift of the total beam-field $\boldsymbol{\mathcal{E}}$ [Eq. (\ref{ER})]. The contribution of the $\mathcal{E}_{0x}^{(I)} \, \hat{\mathbf{x}}^{(I)}$ and $\mathcal{E}_{0y}^{(I)} \, \hat{\mathbf{y}}^{(I)}$ fields in the total input is determined by $\theta_{E}$ [Eq. (\ref{E0x,E0y})] --- which, in the present context of singularity generation, is set to $\theta_{ES}$ determined by Eq. (\ref{tthES0}). This results in a unique variation of GH shift of the total beam-field as a function of $\theta_{i0}$, with the readjustment of $\theta_E$ for every $\theta_{i0}$ value taken into account [Fig. \ref{Fig_thi0xGH}].  
This variation also reveals an unobvious fact that the polarization singularities in the field profiles of Fig. \ref{Fig_EProfilesTh} are not in general formed at the beam-centroid positions; because, the condition of Eq. (\ref{tthES0}) ensures the formation of the singularities at the central point $O$, whereas, the beam-centroid positions are GH-shifted according to the variation shown in Fig. \ref{Fig_thi0xGH}. The variation also shows particularly large (comparable to the effective half-beam-width $w_R$ [Eq. (\ref{GR_def})]) GH shifts for near-Brewster angles of incidence, which is consistent with the effects described by Chan and Tamir \cite{ChanCC}.

Unlike the GH shift, that does not depend on the phase $\Phi_E$ [Eq. (\ref{E0I})], the IF shift depends on all the three parameters $(\theta_{i0},\theta_E,\Phi_E)$. 
As explained graphically in Ref. \cite{ADNKVBrew2021}, the output fields corresponding to the $\mathcal{E}_{0x}^{(I)} \, \hat{\mathbf{x}}^{(I)}$ and $\mathcal{E}_{0y}^{(I)} \, \hat{\mathbf{y}}^{(I)}$ inputs, for a general $\Phi_E$, superpose to create an asymmetry in the total beam-field with respect to the $x$ axis. Because of this asymmetry, the centroid of the total beam-field undergoes a transverse (i.e. along the $y$ axis) shift --- which is the IF shift of the total beam-field. However, we have verified that the assignment $\Phi_E = \Phi_{ES} = \pm \pi/2$ [Eq. (\ref{Condition})] eliminates this asymmetry in the total intensity profile in the present case. The $y$-shift of the total beam-centroid thus reduces to zero, resulting in a zero IF shift in the presently considered case.

\subsection{Spin Shifts and Spin-Hall Effect}

In terms of the presently available complete field information, the spin-shifts are interpreted straightforwardly as the centroid shifts of the $\hat{\boldsymbol{\sigma}}^\pm$ field intensity profiles.
These centroid shifts also depend on the three parameters $(\theta_{i0},\theta_E,\Phi_E)$. For the present demonstration we consider the fixed $\theta_{i0} = 45^\circ$ case of Subsection \ref{Subsec_thEVariation} [Fig. \ref{Fig_Ipm}], and show the centroid-shift characteristics by varying $\theta_E$ around the central $\theta_{ES}$ value $17.28^\circ$.

As mentioned in Fig. \ref{Fig_Ipm}, the maximum intensity of the total field for $(\theta_{i0},\theta_E) = (45^\circ,17.28^\circ)$ is $I_{max} = 4.83$ W/m$^2$, whereas that of the $\hat{\boldsymbol{\sigma}}^+$ field for the same parameter values is $I_{+max} = 0.74$ mW/m$^2$. The dominant $\hat{\boldsymbol{\sigma}}^-$ field thus hugely outweighs the remnant $\hat{\boldsymbol{\sigma}}^+$ field; because of which, the $\hat{\boldsymbol{\sigma}}^-$ field characteristics are near-identical to the total beam-field characteristics. In particular, the longitudinal and transverse shifts of the $\hat{\boldsymbol{\sigma}}^-$ intensity-centroid are respectively near-identical to the GH and IF shifts of the total field --- a special behaviour of the presently considered polarization-singular field.

\begin{figure}
\includegraphics[width = 0.9\linewidth]{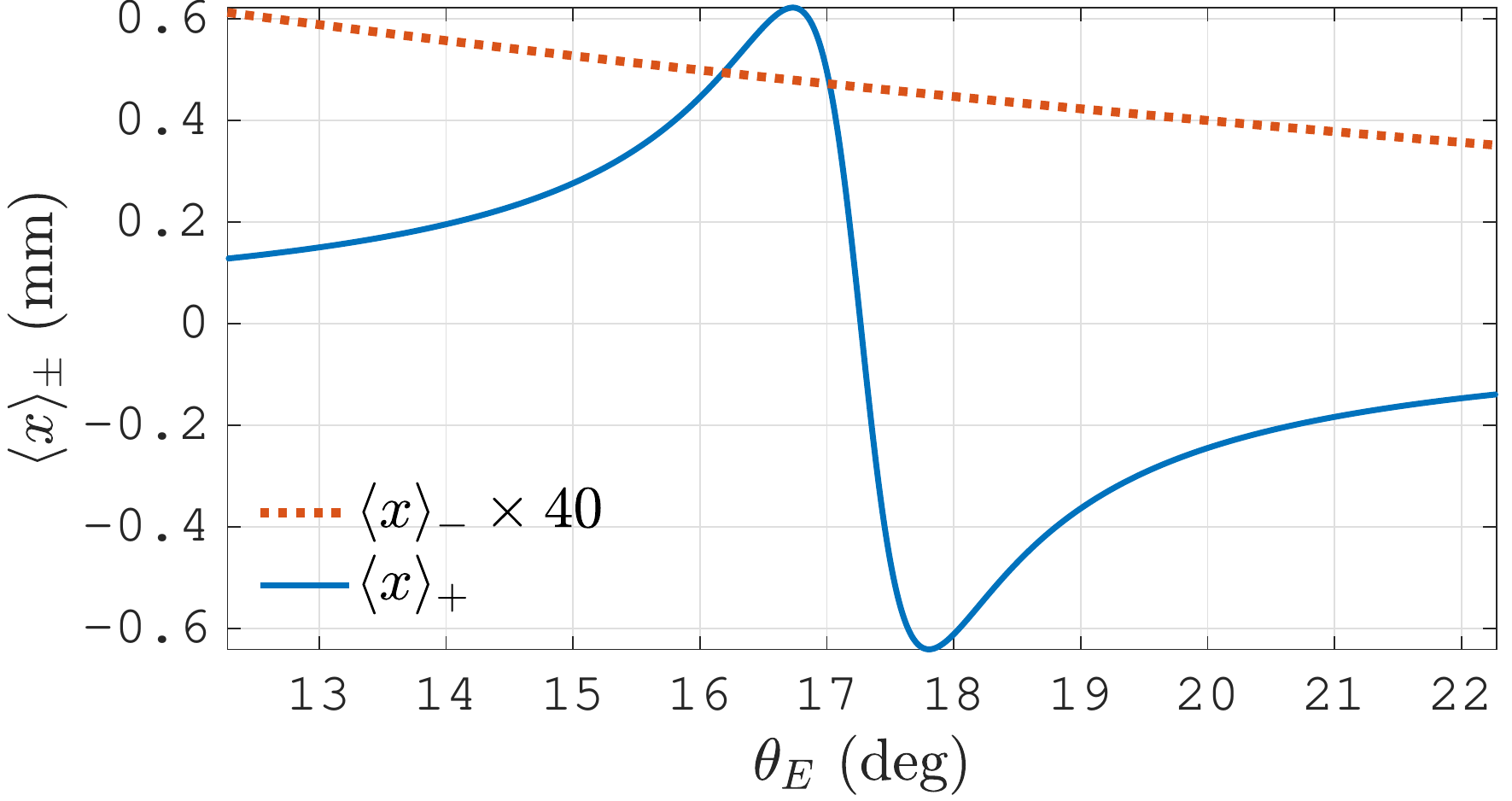}
\caption{Variations of the longitudinal centroid shifts $\langle x\rangle_{\pm}$ of the output $\hat{\boldsymbol{\sigma}}^\pm$ intensity profiles as functions of $\theta_{E}$ around the central $\theta_{ES}$ value $17.28^\circ$ for $\theta_{i0} = 45^\circ$ [Eq. (\ref{tthES0})]. The plot shows large $\hat{\boldsymbol{\sigma}}^+$ spin shifts (comparable to the half-beam-width $w_R$) on both sides of $\theta_{E} = 17.28^\circ$. A zero centroid shift is observed for $\theta_{E} = 17.27^\circ$ but not for $17.28^\circ$, due to the slight asymmetry of the $\hat{\boldsymbol{\sigma}}^+$ intensity profile [Fig. \ref{Fig_Ipm}(e)] with respect to the $y$ axis. The $\hat{\boldsymbol{\sigma}}^-$ spin shift has much smaller values as compared to the $\hat{\boldsymbol{\sigma}}^+$ spin shift; and hence it is represented here by multiplying with 40.}
{\color{grey}{\rule{\linewidth}{1pt}}}
\label{Fig_thExPM}
\end{figure}

The variation of the longitudinal shift of the $\hat{\boldsymbol{\sigma}}^-$ intensity-centroid in a $\theta_E$ range $\pm 5^\circ$ around the central value $17.28^\circ$ is shown in Fig. \ref{Fig_thExPM}. As $\theta_E$ is varied, the contribution of the $\mathcal{E}_{0x}^{(I)} \, \hat{\mathbf{x}}^{(I)}$ and $\mathcal{E}_{0y}^{(I)} \, \hat{\mathbf{y}}^{(I)}$ fields in the total input varies [Eq. (\ref{E0x,E0y})] --- resulting in a variation of the GH shift of the total field and a consequent variation of the longitudinal $\hat{\boldsymbol{\sigma}}^-$ spin-shift. However, due to the consideration of $\Phi_E = \pm\pi/2$, we obtain a $\hat{\boldsymbol{\sigma}}^-$ intensity profile which is symmetric with respect to the $x$ axis. This results in a zero transverse shift of the $\hat{\boldsymbol{\sigma}}^-$ intensity-centroid, which is consistent with the zero IF shift of the total field.

The transverse shift of the $\hat{\boldsymbol{\sigma}}^+$ intensity-centroid is zero likewise. However, due to the transitional dynamics of the phase singularity of the $\hat{\boldsymbol{\sigma}}^+$ field [Figs. \ref{Fig_Ipm}(d), \ref{Fig_Ipm}(e), \ref{Fig_Ipm}(f)], a significant variation of the longitudinal ($x$) position of the $\hat{\boldsymbol{\sigma}}^+$ intensity-centroid is observed. As seen in Fig. \ref{Fig_Ipm}, as the phase singularity moves from the $x<0$ region to the $x>0$ region, the centroid of the $I_+$ profile shifts drastically along the opposite direction --- resulting in giant longitudinal centroid shifts on both sides of $x = 0$, comparable to the effective half-width $w_R$ [Eq. (\ref{GR_def})] of the total field.  
The variation of the longitudinal centroid-shift of the $I_+$ intensity profile is shown in Fig. \ref{Fig_thExPM} for a $\theta_E$ range $\pm 5^\circ$ around the central value $17.28^\circ$ --- where the unique nature of this longitudinal $\hat{\boldsymbol{\sigma}}^+$ spin-shift is clearly observed.

Finally, since the longitudinal shifts of the $\hat{\boldsymbol{\sigma}}^\pm$ intensity-centroids are different, a well-defined longitudinal spin separation is obtained in the present case. However, since the transverse shifts of these centroids are both zero, no corresponding transverse separation is obtained.
The spin-Hall effect of light, which signifies the transverse spin-separation in the beam-field, is thus zero in the presently considered polarization-singular fields. 
We have verified in the simulation that, if we move away from the singularity formation condition of Eq. (\ref{Condition}) by taking $\Phi_E \neq \pm \pi/2$, we obtain $\hat{\boldsymbol{\sigma}}^\pm$ intensity profiles which are asymmetric with respect to the $x$ axis. This results in non-zero transverse shifts of the $\hat{\boldsymbol{\sigma}}^\pm$ intensity-centroids. Except for two special cases $\Phi_E = 0, \pi$, these centroid shifts are unequal, resulting in a transverse separation between the centroids. Non-zero spin-Hall shifts are thus achieved in such cases by perturbing the singularity formation.

\subsection{Further Generalization and Future Directions}

In Subsection \ref{Subsec_PhaseVortex} we have interpreted the manifestation of SOI by identifying the partial conversion from spin to orbital angular momentum. Due to the availability of the complete field information, the next step in the present formalism would be to mathematically characterize this phenomenon by determining the exact spin and orbital contributions to the total angular momentum of the field --- which can be achieved by following the methods of Allen, Beijersbergen, Spreeuw and Woerdman \cite{AllenOAM1992}, Berry \cite{Berry1998} and Barnett \cite{Barnett2022}. The relevant detailed mathematical analysis is outside the scope of the present paper. But this anticipation further strengthens our assertion that all the novel beam-field phenomena are manifestations of different characteristics of the same inhomogeneously-polarized beam-field --- and with the complete information on the beam-field, all these novel phenomena and their interrelations can be seamlessly explained under the same formalism. We strongly anticipate that the complete mathematical characterization of SOI would be particularly significant and fundamentally interesting in the context of the generic optical singular fields considered in the present paper.

Even though we have demonstrated our central context of singularity generation by considering the interrelation of phase and polarization singularities, we emphasize that a phase singularity can exist without an associated polarization singularity. For example, if the optical system is configured to give a certain elliptical polarization $\hat{\mathbf{e}}_1$ in an ideal reflected plane-wave field, the reflected paraxial beam-field attains the polarization $\hat{\mathbf{e}}_1$ only at the beam-center, surrounded by other elliptical polarizations. Clearly, this inhomogeneously polarized beam-field can be interpreted as a superposition of (1) the dominant $\hat{\mathbf{e}}_1$-polarized field with a constant or almost-constant phase profile, and (2) the remnant orthogonally polarized (say $\hat{\mathbf{e}}_2$, orthogonal to $\hat{\mathbf{e}}_1$ on the Poincar\'e sphere) 
field with a phase singularity. Thus, a phase-singular $\hat{\mathbf{e}}_2$-polarized field is obtained, even though the total field does not contain a polarization singularity. Interpreted from an experimental perspective, one can exactly cancel the elliptical polarization at a specific point in the beam-field by using an appropriately configured QWP-GTP combination. This operation removes the contribution of that specific elliptical polarization from the entire beam-field. The controlled output thus obtained contains contribution from only the orthogonal elliptical polarization, with a phase singularity at the point where the original field has been exactly cancelled.

Clearly, any general inhomogeneously-polarized paraxial beam-field can be expressed in this way as a dominant-remnant field superposition. Since the beam shifts, spin shifts and SOI phenomena are simply different manifestations of the fundamental inhomogeneity of the beam-field, the above discussion indicates towards the significant possibility that all such special phenomena arising in a complex reflected (or transmitted) beam-field must be strongly correlatable to the superposition characteristics of a dominant plane-wave/near-plane-wave field and a remnant orthogonally-polarized phase-singular field. 
This possible perspective is a fundamentally appealing generalized view of our present formalism, which can potentially lead to a new direction for future exploration.

Finally, we emphasize that our method of generic singularity formation can find potential application in experimental measurement of refractive indices of dielectric materials. As seen in Fig. \ref{Fig_Ipm}, the singularity position in the beam-field is very sensitive to $\theta_E$. The central $\theta_{ES}$ value, as determined by Eq. (\ref{tthES0}), depends on $\theta_{i0}$ as well as on the refractive indices $n_1$ and $n_2$. Experimentally determined $(\theta_{i0},\theta_{E})$ pairs that create the central singularity as in Fig. \ref{Fig_Ipm}(h) thus contain information on $n_1$ and $n_2$ with significant accuracy. This measurement can thus be utilized to determine an unknown refractive index $n_2$ (while using $n_1 = 1$ for air) --- a potential technique whose efficiency and accuracy is to be explored in the future.


\section{Conclusion} \label{Sec_Conc}

In the present paper, we have identified a generalized condition on the input polarization that can generate a $C$-point polarization singularity in a reflected paraxial beam-field for any central angle of incidence. Associated to this polarization singularity, we have characterized the phase singularity of the constituent field whose polarization is orthogonal to the central circular polarization. Singularity generation due to Brewster reflection \cite{VortexBrewster, CLEO2020, ADNKVBrew2021} is seamlessly understood as a special case of the present formalism. We have demonstrated these polarization and phase singularities via simulated profiles, and have verified them experimentally. Finally, we have demonstrated the dynamics of the singularities due to controlled variation of the input polarization.
The result that phase and polarization singularities can be generated by a single reflection of a paraxial beam at a plane isotropic dielectric interface --- due to any general central angle of incidence --- is the central feature of our formalism.

Even though we have derived the singularity generation condition by using a Fresnel-coefficient-based calculation only at the central plane of incidence, we have demonstrated the simulated field profiles by obtaining exact output field information via our generalized reflection and transmission coefficient matrix formalism. By virtue of this availability of complete field information, we have explored the unique natures of the GH, IF and spin shifts of the optical-singular beam-fields. We have proposed that the SOI phenomena of the presently considered fields can be completely characterized mathematically by using the available exact field information; which would be fundamentally significant to explore especially due to the involvement of optical singularity formations. Finally, by explaining the generation of a phase-singular field independent of a polarization singularity, we have demonstrated a special decomposition of an inhomogeneously polarized field as a superposition of a dominant plane-wave/near-plane-wave field and a remnant orthogonally polarized phase-singular field. The special beam-field phenomena studied in the literature and the optical-singular phenomena discussed in the present paper are all interrelated, as they are manifestations of the same fundamental polarization inhomogeneity of the beam-field. While some of these interrelation are explored in the present paper, we anticipate that further detailed exploration in this direction --- especially a complete mathematical characterization of SOI phenomena in the present singularity formation context --- would provide a very rich and significant understanding on the fundamental polarization-inhomogeneity characteristics of a reflected paraxial beam-field.
Additionally, our singularity formation method can have potential application in experimental characterization of dielectric media, especially in measurements of unknown refractive indices.

\begin{acknowledgments}
A.D. thanks CSIR for Senior Research Fellowship. N.K.V. thanks SERB for financial support.
\end{acknowledgments}


\bibliography{AD_NKV_GenVortex_Refs}

\begin{thebibliography}{62}%
\makeatletter
\providecommand \@ifxundefined [1]{%
 \@ifx{#1\undefined}
}%
\providecommand \@ifnum [1]{%
 \ifnum #1\expandafter \@firstoftwo
 \else \expandafter \@secondoftwo
 \fi
}%
\providecommand \@ifx [1]{%
 \ifx #1\expandafter \@firstoftwo
 \else \expandafter \@secondoftwo
 \fi
}%
\providecommand \natexlab [1]{#1}%
\providecommand \enquote  [1]{``#1''}%
\providecommand \bibnamefont  [1]{#1}%
\providecommand \bibfnamefont [1]{#1}%
\providecommand \citenamefont [1]{#1}%
\providecommand \href@noop [0]{\@secondoftwo}%
\providecommand \href [0]{\begingroup \@sanitize@url \@href}%
\providecommand \@href[1]{\@@startlink{#1}\@@href}%
\providecommand \@@href[1]{\endgroup#1\@@endlink}%
\providecommand \@sanitize@url [0]{\catcode `\\12\catcode `\$12\catcode
  `\&12\catcode `\#12\catcode `\^12\catcode `\_12\catcode `\%12\relax}%
\providecommand \@@startlink[1]{}%
\providecommand \@@endlink[0]{}%
\providecommand \url  [0]{\begingroup\@sanitize@url \@url }%
\providecommand \@url [1]{\endgroup\@href {#1}{\urlprefix }}%
\providecommand \urlprefix  [0]{URL }%
\providecommand \Eprint [0]{\href }%
\providecommand \doibase [0]{https://doi.org/}%
\providecommand \selectlanguage [0]{\@gobble}%
\providecommand \bibinfo  [0]{\@secondoftwo}%
\providecommand \bibfield  [0]{\@secondoftwo}%
\providecommand \translation [1]{[#1]}%
\providecommand \BibitemOpen [0]{}%
\providecommand \bibitemStop [0]{}%
\providecommand \bibitemNoStop [0]{.\EOS\space}%
\providecommand \EOS [0]{\spacefactor3000\relax}%
\providecommand \BibitemShut  [1]{\csname bibitem#1\endcsname}%
\let\auto@bib@innerbib\@empty
\bibitem [{\citenamefont {Born}\ and\ \citenamefont {Wolf}(1999)}]{BornWolf}%
  \BibitemOpen
  \bibfield  {author} {\bibinfo {author} {\bibfnamefont {M.}~\bibnamefont
  {Born}}\ and\ \bibinfo {author} {\bibfnamefont {E.}~\bibnamefont {Wolf}},\
  }\href@noop {} {\emph {\bibinfo {title} {Principles of Optics}}},\ \bibinfo
  {edition} {7th}\ ed.\ (\bibinfo  {publisher} {Cambridge University Press,
  Cambridge},\ \bibinfo {year} {1999})\BibitemShut {NoStop}%
\bibitem [{\citenamefont {Goos}\ and\ \citenamefont {H\"anchen}(1947)}]{GH}%
  \BibitemOpen
  \bibfield  {author} {\bibinfo {author} {\bibfnamefont {V.~F.}\ \bibnamefont
  {Goos}}\ and\ \bibinfo {author} {\bibfnamefont {H.}~\bibnamefont
  {H\"anchen}},\ }\href@noop {} {\bibfield  {journal} {\bibinfo  {journal}
  {Ann. Physik}\ }\textbf {\bibinfo {volume} {436}},\ \bibinfo {pages} {333}
  (\bibinfo {year} {1947})}\BibitemShut {NoStop}%
\bibitem [{\citenamefont {Artmann}(1948)}]{Artmann}%
  \BibitemOpen
  \bibfield  {author} {\bibinfo {author} {\bibfnamefont {K.}~\bibnamefont
  {Artmann}},\ }\href@noop {} {\bibfield  {journal} {\bibinfo  {journal} {Ann.
  Physik}\ }\textbf {\bibinfo {volume} {437}},\ \bibinfo {pages} {87} (\bibinfo
  {year} {1948})}\BibitemShut {NoStop}%
\bibitem [{\citenamefont {Ra}\ \emph {et~al.}(1973)\citenamefont {Ra},
  \citenamefont {Bertoni},\ and\ \citenamefont {Felsen}}]{RaJW}%
  \BibitemOpen
  \bibfield  {author} {\bibinfo {author} {\bibfnamefont {J.~W.}\ \bibnamefont
  {Ra}}, \bibinfo {author} {\bibfnamefont {H.~L.}\ \bibnamefont {Bertoni}},\
  and\ \bibinfo {author} {\bibfnamefont {L.~B.}\ \bibnamefont {Felsen}},\
  }\href@noop {} {\bibfield  {journal} {\bibinfo  {journal} {SIAM J. Appl.
  Math.}\ }\textbf {\bibinfo {volume} {24}},\ \bibinfo {pages} {396} (\bibinfo
  {year} {1973})}\BibitemShut {NoStop}%
\bibitem [{\citenamefont {Antar}\ and\ \citenamefont
  {Boerner}(1974)}]{AntarYM}%
  \BibitemOpen
  \bibfield  {author} {\bibinfo {author} {\bibfnamefont {Y.~M.}\ \bibnamefont
  {Antar}}\ and\ \bibinfo {author} {\bibfnamefont {W.~M.}\ \bibnamefont
  {Boerner}},\ }\href@noop {} {\bibfield  {journal} {\bibinfo  {journal} {Can.
  J. Phys.}\ }\textbf {\bibinfo {volume} {52}},\ \bibinfo {pages} {962}
  (\bibinfo {year} {1974})}\BibitemShut {NoStop}%
\bibitem [{\citenamefont {McGuirk}\ and\ \citenamefont
  {Carniglia}(1977)}]{McGuirk}%
  \BibitemOpen
  \bibfield  {author} {\bibinfo {author} {\bibfnamefont {M.}~\bibnamefont
  {McGuirk}}\ and\ \bibinfo {author} {\bibfnamefont {C.~K.}\ \bibnamefont
  {Carniglia}},\ }\href@noop {} {\bibfield  {journal} {\bibinfo  {journal} {J.
  Opt. Soc. Am.}\ }\textbf {\bibinfo {volume} {67}},\ \bibinfo {pages} {103}
  (\bibinfo {year} {1977})}\BibitemShut {NoStop}%
\bibitem [{\citenamefont {Chan}\ and\ \citenamefont {Tamir}(1985)}]{ChanCC}%
  \BibitemOpen
  \bibfield  {author} {\bibinfo {author} {\bibfnamefont {C.~C.}\ \bibnamefont
  {Chan}}\ and\ \bibinfo {author} {\bibfnamefont {C.}~\bibnamefont {Tamir}},\
  }\href@noop {} {\bibfield  {journal} {\bibinfo  {journal} {Opt. Lett.}\
  }\textbf {\bibinfo {volume} {10}},\ \bibinfo {pages} {378} (\bibinfo {year}
  {1985})}\BibitemShut {NoStop}%
\bibitem [{\citenamefont {Porras}(1996)}]{Porras}%
  \BibitemOpen
  \bibfield  {author} {\bibinfo {author} {\bibfnamefont {M.~A.}\ \bibnamefont
  {Porras}},\ }\href@noop {} {\bibfield  {journal} {\bibinfo  {journal} {Optics
  Communications}\ }\textbf {\bibinfo {volume} {131}},\ \bibinfo {pages} {13}
  (\bibinfo {year} {1996})}\BibitemShut {NoStop}%
\bibitem [{\citenamefont {Aiello}\ and\ \citenamefont
  {Woerdman}(2009)}]{AielloArXiv}%
  \BibitemOpen
  \bibfield  {author} {\bibinfo {author} {\bibfnamefont {A.}~\bibnamefont
  {Aiello}}\ and\ \bibinfo {author} {\bibfnamefont {J.~P.}\ \bibnamefont
  {Woerdman}},\ }\href@noop {} {\bibfield  {journal} {\bibinfo  {journal}
  {arXiv:0903.3730v2 [physics.optics]}\ } (\bibinfo {year} {2009})}\BibitemShut
  {NoStop}%
\bibitem [{\citenamefont {Fedorov}(1955)}]{Fedorov}%
  \BibitemOpen
  \bibfield  {author} {\bibinfo {author} {\bibfnamefont {F.~I.}\ \bibnamefont
  {Fedorov}},\ }\href@noop {} {\bibfield  {journal} {\bibinfo  {journal} {Dokl.
  Akad. Nauk SSSR}\ }\textbf {\bibinfo {volume} {105}},\ \bibinfo {pages} {465}
  (\bibinfo {year} {1955})},\ \bibinfo {note} {english translation available at
  http://master. basnet.by/congress2011/symposium/spbi.pdf}\BibitemShut
  {NoStop}%
\bibitem [{\citenamefont {Schilling}(1965)}]{Schilling}%
  \BibitemOpen
  \bibfield  {author} {\bibinfo {author} {\bibfnamefont {H.}~\bibnamefont
  {Schilling}},\ }\href@noop {} {\bibfield  {journal} {\bibinfo  {journal}
  {Ann. Physik}\ }\textbf {\bibinfo {volume} {471}},\ \bibinfo {pages} {122}
  (\bibinfo {year} {1965})}\BibitemShut {NoStop}%
\bibitem [{\citenamefont {Imbert}(1972)}]{Imbert}%
  \BibitemOpen
  \bibfield  {author} {\bibinfo {author} {\bibfnamefont {C.}~\bibnamefont
  {Imbert}},\ }\href@noop {} {\bibfield  {journal} {\bibinfo  {journal} {Phys.
  Rev. D}\ }\textbf {\bibinfo {volume} {5}},\ \bibinfo {pages} {787} (\bibinfo
  {year} {1972})}\BibitemShut {NoStop}%
\bibitem [{\citenamefont {Player}(1987)}]{Player}%
  \BibitemOpen
  \bibfield  {author} {\bibinfo {author} {\bibfnamefont {M.~A.}\ \bibnamefont
  {Player}},\ }\href@noop {} {\bibfield  {journal} {\bibinfo  {journal} {J.
  Phys. A: Math. Gen.}\ }\textbf {\bibinfo {volume} {20}},\ \bibinfo {pages}
  {3667} (\bibinfo {year} {1987})}\BibitemShut {NoStop}%
\bibitem [{\citenamefont {Fedoseyev}(1988)}]{FVG}%
  \BibitemOpen
  \bibfield  {author} {\bibinfo {author} {\bibfnamefont {V.~G.}\ \bibnamefont
  {Fedoseyev}},\ }\href@noop {} {\bibfield  {journal} {\bibinfo  {journal} {J.
  Phys. A: Math. Gen.}\ }\textbf {\bibinfo {volume} {21}},\ \bibinfo {pages}
  {2045} (\bibinfo {year} {1988})}\BibitemShut {NoStop}%
\bibitem [{\citenamefont {Liberman}\ and\ \citenamefont
  {Zel’dovich}(1992)}]{Liberman}%
  \BibitemOpen
  \bibfield  {author} {\bibinfo {author} {\bibfnamefont {V.~S.}\ \bibnamefont
  {Liberman}}\ and\ \bibinfo {author} {\bibfnamefont {B.~Y.}\ \bibnamefont
  {Zel’dovich}},\ }\href@noop {} {\bibfield  {journal} {\bibinfo  {journal}
  {Phys. Rev. A}\ }\textbf {\bibinfo {volume} {46}},\ \bibinfo {pages} {5199}
  (\bibinfo {year} {1992})}\BibitemShut {NoStop}%
\bibitem [{\citenamefont {Onoda}\ \emph {et~al.}(2004)\citenamefont {Onoda},
  \citenamefont {Murakami},\ and\ \citenamefont {Nagaosa}}]{Onoda}%
  \BibitemOpen
  \bibfield  {author} {\bibinfo {author} {\bibfnamefont {M.}~\bibnamefont
  {Onoda}}, \bibinfo {author} {\bibfnamefont {S.}~\bibnamefont {Murakami}},\
  and\ \bibinfo {author} {\bibfnamefont {N.}~\bibnamefont {Nagaosa}},\
  }\href@noop {} {\bibfield  {journal} {\bibinfo  {journal} {Phys. Rev. Lett.}\
  }\textbf {\bibinfo {volume} {93}},\ \bibinfo {pages} {083901} (\bibinfo
  {year} {2004})}\BibitemShut {NoStop}%
\bibitem [{\citenamefont {Bliokh}\ and\ \citenamefont
  {Bliokh}(2006)}]{Bliokh2006}%
  \BibitemOpen
  \bibfield  {author} {\bibinfo {author} {\bibfnamefont {K.~Y.}\ \bibnamefont
  {Bliokh}}\ and\ \bibinfo {author} {\bibfnamefont {Y.~P.}\ \bibnamefont
  {Bliokh}},\ }\href@noop {} {\bibfield  {journal} {\bibinfo  {journal} {Phys.
  Rev. Lett.}\ }\textbf {\bibinfo {volume} {96}},\ \bibinfo {pages} {073903}
  (\bibinfo {year} {2006})}\BibitemShut {NoStop}%
\bibitem [{\citenamefont {Bliokh}\ and\ \citenamefont
  {Bliokh}(2007)}]{Bliokh2007}%
  \BibitemOpen
  \bibfield  {author} {\bibinfo {author} {\bibfnamefont {K.~Y.}\ \bibnamefont
  {Bliokh}}\ and\ \bibinfo {author} {\bibfnamefont {Y.~P.}\ \bibnamefont
  {Bliokh}},\ }\href@noop {} {\bibfield  {journal} {\bibinfo  {journal} {Phys.
  Rev. E}\ }\textbf {\bibinfo {volume} {75}},\ \bibinfo {pages} {066609}
  (\bibinfo {year} {2007})}\BibitemShut {NoStop}%
\bibitem [{\citenamefont {Hosten}\ and\ \citenamefont
  {Kwiat}(2008)}]{HostenKwiat}%
  \BibitemOpen
  \bibfield  {author} {\bibinfo {author} {\bibfnamefont {O.}~\bibnamefont
  {Hosten}}\ and\ \bibinfo {author} {\bibfnamefont {P.}~\bibnamefont {Kwiat}},\
  }\href@noop {} {\bibfield  {journal} {\bibinfo  {journal} {Science}\ }\textbf
  {\bibinfo {volume} {319}},\ \bibinfo {pages} {787} (\bibinfo {year}
  {2008})}\BibitemShut {NoStop}%
\bibitem [{\citenamefont {Aiello}\ and\ \citenamefont
  {Woerdman}(2007)}]{AielloArXiv2}%
  \BibitemOpen
  \bibfield  {author} {\bibinfo {author} {\bibfnamefont {A.}~\bibnamefont
  {Aiello}}\ and\ \bibinfo {author} {\bibfnamefont {J.~P.}\ \bibnamefont
  {Woerdman}},\ }\href@noop {} {\bibfield  {journal} {\bibinfo  {journal}
  {arXiv:0710.1643v2 [physics.optics]}\ } (\bibinfo {year} {2007})}\BibitemShut
  {NoStop}%
\bibitem [{\citenamefont {Aiello}\ and\ \citenamefont
  {Woerdman}(2008)}]{Aiello2008}%
  \BibitemOpen
  \bibfield  {author} {\bibinfo {author} {\bibfnamefont {A.}~\bibnamefont
  {Aiello}}\ and\ \bibinfo {author} {\bibfnamefont {J.~P.}\ \bibnamefont
  {Woerdman}},\ }\href@noop {} {\bibfield  {journal} {\bibinfo  {journal} {Opt.
  Lett.}\ }\textbf {\bibinfo {volume} {33}},\ \bibinfo {pages} {1437} (\bibinfo
  {year} {2008})}\BibitemShut {NoStop}%
\bibitem [{\citenamefont {Merano}\ \emph {et~al.}(2009)\citenamefont {Merano},
  \citenamefont {Aiello}, \citenamefont {van Exter},\ and\ \citenamefont
  {Woerdman}}]{Merano2009}%
  \BibitemOpen
  \bibfield  {author} {\bibinfo {author} {\bibfnamefont {M.}~\bibnamefont
  {Merano}}, \bibinfo {author} {\bibfnamefont {A.}~\bibnamefont {Aiello}},
  \bibinfo {author} {\bibfnamefont {M.~P.}\ \bibnamefont {van Exter}},\ and\
  \bibinfo {author} {\bibfnamefont {J.~P.}\ \bibnamefont {Woerdman}},\
  }\href@noop {} {\bibfield  {journal} {\bibinfo  {journal} {Nature Photon}\
  }\textbf {\bibinfo {volume} {3}},\ \bibinfo {pages} {337} (\bibinfo {year}
  {2009})}\BibitemShut {NoStop}%
\bibitem [{\citenamefont {Aiello}\ \emph {et~al.}(2009)\citenamefont {Aiello},
  \citenamefont {Merano},\ and\ \citenamefont {Woerdman}}]{Aiello2009}%
  \BibitemOpen
  \bibfield  {author} {\bibinfo {author} {\bibfnamefont {A.}~\bibnamefont
  {Aiello}}, \bibinfo {author} {\bibfnamefont {M.}~\bibnamefont {Merano}},\
  and\ \bibinfo {author} {\bibfnamefont {J.~P.}\ \bibnamefont {Woerdman}},\
  }\href@noop {} {\bibfield  {journal} {\bibinfo  {journal} {Phys. Rev. A}\
  }\textbf {\bibinfo {volume} {80}},\ \bibinfo {pages} {061801(R)} (\bibinfo
  {year} {2009})}\BibitemShut {NoStop}%
\bibitem [{\citenamefont {Qin}\ \emph {et~al.}(2011)\citenamefont {Qin},
  \citenamefont {Li}, \citenamefont {Feng}, \citenamefont {Xiao}, \citenamefont
  {Yang},\ and\ \citenamefont {Gong}}]{Qin2011}%
  \BibitemOpen
  \bibfield  {author} {\bibinfo {author} {\bibfnamefont {Y.}~\bibnamefont
  {Qin}}, \bibinfo {author} {\bibfnamefont {Y.}~\bibnamefont {Li}}, \bibinfo
  {author} {\bibfnamefont {X.}~\bibnamefont {Feng}}, \bibinfo {author}
  {\bibfnamefont {Y.-F.}\ \bibnamefont {Xiao}}, \bibinfo {author}
  {\bibfnamefont {H.}~\bibnamefont {Yang}},\ and\ \bibinfo {author}
  {\bibfnamefont {Q.}~\bibnamefont {Gong}},\ }\href@noop {} {\bibfield
  {journal} {\bibinfo  {journal} {Opt. Express}\ }\textbf {\bibinfo {volume}
  {19}},\ \bibinfo {pages} {9636} (\bibinfo {year} {2011})}\BibitemShut
  {NoStop}%
\bibitem [{\citenamefont {Bliokh}\ and\ \citenamefont {Aiello}(2013)}]{BARev}%
  \BibitemOpen
  \bibfield  {author} {\bibinfo {author} {\bibfnamefont {K.~Y.}\ \bibnamefont
  {Bliokh}}\ and\ \bibinfo {author} {\bibfnamefont {A.}~\bibnamefont
  {Aiello}},\ }\href@noop {} {\bibfield  {journal} {\bibinfo  {journal} {J.
  Opt.}\ }\textbf {\bibinfo {volume} {15}},\ \bibinfo {pages} {014001}
  (\bibinfo {year} {2013})}\BibitemShut {NoStop}%
\bibitem [{\citenamefont {G\"otte}\ \emph {et~al.}(2014)\citenamefont
  {G\"otte}, \citenamefont {L\"offler},\ and\ \citenamefont
  {Dennis}}]{GotteLofflerDennis}%
  \BibitemOpen
  \bibfield  {author} {\bibinfo {author} {\bibfnamefont {J.~B.}\ \bibnamefont
  {G\"otte}}, \bibinfo {author} {\bibfnamefont {W.}~\bibnamefont {L\"offler}},\
  and\ \bibinfo {author} {\bibfnamefont {M.~R.}\ \bibnamefont {Dennis}},\
  }\href@noop {} {\bibfield  {journal} {\bibinfo  {journal} {Phys. Rev. Lett.}\
  }\textbf {\bibinfo {volume} {112}},\ \bibinfo {pages} {233901} (\bibinfo
  {year} {2014})}\BibitemShut {NoStop}%
\bibitem [{\citenamefont {Xie}\ \emph {et~al.}(2018)\citenamefont {Xie},
  \citenamefont {Zhou}, \citenamefont {Qiu}, \citenamefont {Luo}, \citenamefont
  {Liu}, \citenamefont {Li}, \citenamefont {He}, \citenamefont {Du},
  \citenamefont {Zhang},\ and\ \citenamefont {Wang}}]{XieSHELinIF}%
  \BibitemOpen
  \bibfield  {author} {\bibinfo {author} {\bibfnamefont {L.}~\bibnamefont
  {Xie}}, \bibinfo {author} {\bibfnamefont {X.}~\bibnamefont {Zhou}}, \bibinfo
  {author} {\bibfnamefont {X.}~\bibnamefont {Qiu}}, \bibinfo {author}
  {\bibfnamefont {L.}~\bibnamefont {Luo}}, \bibinfo {author} {\bibfnamefont
  {X.}~\bibnamefont {Liu}}, \bibinfo {author} {\bibfnamefont {Z.}~\bibnamefont
  {Li}}, \bibinfo {author} {\bibfnamefont {Y.}~\bibnamefont {He}}, \bibinfo
  {author} {\bibfnamefont {J.}~\bibnamefont {Du}}, \bibinfo {author}
  {\bibfnamefont {Z.}~\bibnamefont {Zhang}},\ and\ \bibinfo {author}
  {\bibfnamefont {D.}~\bibnamefont {Wang}},\ }\href@noop {} {\bibfield
  {journal} {\bibinfo  {journal} {Opt. Express}\ }\textbf {\bibinfo {volume}
  {26}},\ \bibinfo {pages} {22934} (\bibinfo {year} {2018})}\BibitemShut
  {NoStop}%
\bibitem [{\citenamefont {Gbur}(2017)}]{Gbur}%
  \BibitemOpen
  \bibfield  {author} {\bibinfo {author} {\bibfnamefont {G.~J.}\ \bibnamefont
  {Gbur}},\ }\href@noop {} {\emph {\bibinfo {title} {Singular Optics}}}\
  (\bibinfo  {publisher} {CRC Press, Taylor \& Francis Group, LLC},\ \bibinfo
  {address} {FL},\ \bibinfo {year} {2017})\BibitemShut {NoStop}%
\bibitem [{\citenamefont {Nye}\ and\ \citenamefont
  {Berry}(1974)}]{NyeBerry1974}%
  \BibitemOpen
  \bibfield  {author} {\bibinfo {author} {\bibfnamefont {J.~F.}\ \bibnamefont
  {Nye}}\ and\ \bibinfo {author} {\bibfnamefont {M.~V.}\ \bibnamefont
  {Berry}},\ }\href@noop {} {\bibfield  {journal} {\bibinfo  {journal} {Proc.
  R. Soc. Lond. A}\ }\textbf {\bibinfo {volume} {336}},\ \bibinfo {pages} {165}
  (\bibinfo {year} {1974})}\BibitemShut {NoStop}%
\bibitem [{\citenamefont {Bhandari}(1997)}]{Bhandari97}%
  \BibitemOpen
  \bibfield  {author} {\bibinfo {author} {\bibfnamefont {R.}~\bibnamefont
  {Bhandari}},\ }\href@noop {} {\bibfield  {journal} {\bibinfo  {journal}
  {Phys. Rep.}\ }\textbf {\bibinfo {volume} {281}},\ \bibinfo {pages} {1}
  (\bibinfo {year} {1997})}\BibitemShut {NoStop}%
\bibitem [{\citenamefont {Soskin}\ \emph {et~al.}(1997)\citenamefont {Soskin},
  \citenamefont {Gorshkov},\ and\ \citenamefont {Vasnetsov}}]{SGV97}%
  \BibitemOpen
  \bibfield  {author} {\bibinfo {author} {\bibfnamefont {M.~S.}\ \bibnamefont
  {Soskin}}, \bibinfo {author} {\bibfnamefont {V.~N.}\ \bibnamefont
  {Gorshkov}},\ and\ \bibinfo {author} {\bibfnamefont {M.~V.}\ \bibnamefont
  {Vasnetsov}},\ }\href@noop {} {\bibfield  {journal} {\bibinfo  {journal}
  {Phys. Rev. A}\ }\textbf {\bibinfo {volume} {56}},\ \bibinfo {pages} {4064}
  (\bibinfo {year} {1997})}\BibitemShut {NoStop}%
\bibitem [{\citenamefont {Padgett}\ and\ \citenamefont {Allen}(2000)}]{PA2000}%
  \BibitemOpen
  \bibfield  {author} {\bibinfo {author} {\bibfnamefont {M.}~\bibnamefont
  {Padgett}}\ and\ \bibinfo {author} {\bibfnamefont {A.}~\bibnamefont
  {Allen}},\ }\href@noop {} {\bibfield  {journal} {\bibinfo  {journal}
  {Contemp. Phys.}\ }\textbf {\bibinfo {volume} {41}},\ \bibinfo {pages} {275}
  (\bibinfo {year} {2000})}\BibitemShut {NoStop}%
\bibitem [{\citenamefont {Soskin}\ and\ \citenamefont
  {Vasnetsov}(2001)}]{SV2001}%
  \BibitemOpen
  \bibfield  {author} {\bibinfo {author} {\bibfnamefont {M.~S.}\ \bibnamefont
  {Soskin}}\ and\ \bibinfo {author} {\bibfnamefont {M.~V.}\ \bibnamefont
  {Vasnetsov}},\ }\href@noop {} {\bibfield  {journal} {\bibinfo  {journal}
  {Prog. Opt.}\ }\textbf {\bibinfo {volume} {42}},\ \bibinfo {pages} {219}
  (\bibinfo {year} {2001})}\BibitemShut {NoStop}%
\bibitem [{\citenamefont {Dennis}\ \emph {et~al.}(2009)\citenamefont {Dennis},
  \citenamefont {O'Holleran},\ and\ \citenamefont {Padgett}}]{DOP09}%
  \BibitemOpen
  \bibfield  {author} {\bibinfo {author} {\bibfnamefont {M.~R.}\ \bibnamefont
  {Dennis}}, \bibinfo {author} {\bibfnamefont {K.}~\bibnamefont {O'Holleran}},\
  and\ \bibinfo {author} {\bibfnamefont {M.~J.}\ \bibnamefont {Padgett}},\
  }\href@noop {} {\bibfield  {journal} {\bibinfo  {journal} {Prog. Opt.}\
  }\textbf {\bibinfo {volume} {53}},\ \bibinfo {pages} {293} (\bibinfo {year}
  {2009})}\BibitemShut {NoStop}%
\bibitem [{\citenamefont {Bliokh}\ and\ \citenamefont {Nori}(2015)}]{BNRev}%
  \BibitemOpen
  \bibfield  {author} {\bibinfo {author} {\bibfnamefont {K.~Y.}\ \bibnamefont
  {Bliokh}}\ and\ \bibinfo {author} {\bibfnamefont {F.}~\bibnamefont {Nori}},\
  }\href@noop {} {\bibfield  {journal} {\bibinfo  {journal} {Phys. Rep.}\
  }\textbf {\bibinfo {volume} {592}},\ \bibinfo {pages} {1} (\bibinfo {year}
  {2015})}\BibitemShut {NoStop}%
\bibitem [{\citenamefont {Nye}(1983{\natexlab{a}})}]{Nye83b}%
  \BibitemOpen
  \bibfield  {author} {\bibinfo {author} {\bibfnamefont {J.~F.}\ \bibnamefont
  {Nye}},\ }\href@noop {} {\bibfield  {journal} {\bibinfo  {journal} {Proc. R.
  Soc. Lond. A}\ }\textbf {\bibinfo {volume} {387}},\ \bibinfo {pages} {105}
  (\bibinfo {year} {1983}{\natexlab{a}})}\BibitemShut {NoStop}%
\bibitem [{\citenamefont {Nye}(1983{\natexlab{b}})}]{Nye83a}%
  \BibitemOpen
  \bibfield  {author} {\bibinfo {author} {\bibfnamefont {J.~F.}\ \bibnamefont
  {Nye}},\ }\href@noop {} {\bibfield  {journal} {\bibinfo  {journal} {Proc. R.
  Soc. Lond. A}\ }\textbf {\bibinfo {volume} {389}},\ \bibinfo {pages} {279}
  (\bibinfo {year} {1983}{\natexlab{b}})}\BibitemShut {NoStop}%
\bibitem [{\citenamefont {Hajnal}(1987{\natexlab{a}})}]{Hajnal87a}%
  \BibitemOpen
  \bibfield  {author} {\bibinfo {author} {\bibfnamefont {J.~V.}\ \bibnamefont
  {Hajnal}},\ }\href@noop {} {\bibfield  {journal} {\bibinfo  {journal} {Proc.
  R. Soc. Lond. A}\ }\textbf {\bibinfo {volume} {414}},\ \bibinfo {pages} {433}
  (\bibinfo {year} {1987}{\natexlab{a}})}\BibitemShut {NoStop}%
\bibitem [{\citenamefont {Hajnal}(1987{\natexlab{b}})}]{Hajnal87b}%
  \BibitemOpen
  \bibfield  {author} {\bibinfo {author} {\bibfnamefont {J.~V.}\ \bibnamefont
  {Hajnal}},\ }\href@noop {} {\bibfield  {journal} {\bibinfo  {journal} {Proc.
  R. Soc. Lond. A}\ }\textbf {\bibinfo {volume} {414}},\ \bibinfo {pages} {447}
  (\bibinfo {year} {1987}{\natexlab{b}})}\BibitemShut {NoStop}%
\bibitem [{\citenamefont {Nye}\ and\ \citenamefont {Hajnal}(1987)}]{NH1987}%
  \BibitemOpen
  \bibfield  {author} {\bibinfo {author} {\bibfnamefont {J.~F.}\ \bibnamefont
  {Nye}}\ and\ \bibinfo {author} {\bibfnamefont {J.~V.}\ \bibnamefont
  {Hajnal}},\ }\href@noop {} {\bibfield  {journal} {\bibinfo  {journal} {Proc.
  R. Soc. Lond. A}\ }\textbf {\bibinfo {volume} {409}},\ \bibinfo {pages} {21}
  (\bibinfo {year} {1987})}\BibitemShut {NoStop}%
\bibitem [{\citenamefont {Delmarcelle}\ and\ \citenamefont
  {Hesselink}(1994)}]{DH1994}%
  \BibitemOpen
  \bibfield  {author} {\bibinfo {author} {\bibfnamefont {T.}~\bibnamefont
  {Delmarcelle}}\ and\ \bibinfo {author} {\bibfnamefont {L.}~\bibnamefont
  {Hesselink}},\ }\href@noop {} {\bibfield  {journal} {\bibinfo  {journal} {VIS
  ’94: Proceedings of the Conference on Visualization ’94}\ ,\ \bibinfo
  {pages} {140}} (\bibinfo {year} {1994})}\BibitemShut {NoStop}%
\bibitem [{\citenamefont {Dennis}(2002)}]{DennisPS02}%
  \BibitemOpen
  \bibfield  {author} {\bibinfo {author} {\bibfnamefont {M.~R.}\ \bibnamefont
  {Dennis}},\ }\href@noop {} {\bibfield  {journal} {\bibinfo  {journal} {Opt.
  Commun.}\ }\textbf {\bibinfo {volume} {213}},\ \bibinfo {pages} {201}
  (\bibinfo {year} {2002})}\BibitemShut {NoStop}%
\bibitem [{\citenamefont {Dennis}(2008)}]{DennisMonstar08}%
  \BibitemOpen
  \bibfield  {author} {\bibinfo {author} {\bibfnamefont {M.~R.}\ \bibnamefont
  {Dennis}},\ }\href@noop {} {\bibfield  {journal} {\bibinfo  {journal} {Opt.
  Lett.}\ }\textbf {\bibinfo {volume} {33}},\ \bibinfo {pages} {2572} (\bibinfo
  {year} {2008})}\BibitemShut {NoStop}%
\bibitem [{\citenamefont {Viswanathan}\ \emph {et~al.}(2013)\citenamefont
  {Viswanathan}, \citenamefont {Kumar},\ and\ \citenamefont
  {Philip}}]{NKVMonstar}%
  \BibitemOpen
  \bibfield  {author} {\bibinfo {author} {\bibfnamefont {N.~K.}\ \bibnamefont
  {Viswanathan}}, \bibinfo {author} {\bibfnamefont {V.}~\bibnamefont {Kumar}},\
  and\ \bibinfo {author} {\bibfnamefont {G.~M.}\ \bibnamefont {Philip}},\
  }\href@noop {} {\bibfield  {journal} {\bibinfo  {journal} {J. Opt.}\ }\textbf
  {\bibinfo {volume} {15}},\ \bibinfo {pages} {044027} (\bibinfo {year}
  {2013})}\BibitemShut {NoStop}%
\bibitem [{\citenamefont {Jayasurya}\ \emph {et~al.}(2011)\citenamefont
  {Jayasurya}, \citenamefont {Inavalli},\ and\ \citenamefont
  {Viswanathan}}]{NKVFiber}%
  \BibitemOpen
  \bibfield  {author} {\bibinfo {author} {\bibfnamefont {Y.~V.}\ \bibnamefont
  {Jayasurya}}, \bibinfo {author} {\bibfnamefont {V.~V. G.~K.}\ \bibnamefont
  {Inavalli}},\ and\ \bibinfo {author} {\bibfnamefont {N.~K.}\ \bibnamefont
  {Viswanathan}},\ }\href@noop {} {\bibfield  {journal} {\bibinfo  {journal}
  {Appl. Opt.}\ }\textbf {\bibinfo {volume} {50}},\ \bibinfo {pages} {E131}
  (\bibinfo {year} {2011})}\BibitemShut {NoStop}%
\bibitem [{\citenamefont {Ruchi}\ \emph {et~al.}(2017)\citenamefont {Ruchi},
  \citenamefont {Pal},\ and\ \citenamefont {Senthilkumaran}}]{Vpoint}%
  \BibitemOpen
  \bibfield  {author} {\bibinfo {author} {\bibnamefont {Ruchi}}, \bibinfo
  {author} {\bibfnamefont {S.~K.}\ \bibnamefont {Pal}},\ and\ \bibinfo {author}
  {\bibfnamefont {P.}~\bibnamefont {Senthilkumaran}},\ }\href@noop {}
  {\bibfield  {journal} {\bibinfo  {journal} {Opt. Express}\ }\textbf {\bibinfo
  {volume} {25}},\ \bibinfo {pages} {19326} (\bibinfo {year}
  {2017})}\BibitemShut {NoStop}%
\bibitem [{\citenamefont {Barczyk}\ \emph {et~al.}(2019)\citenamefont
  {Barczyk}, \citenamefont {Nechayev}, \citenamefont {Butt}, \citenamefont
  {Leuchs},\ and\ \citenamefont {Banzer}}]{VortexBrewster}%
  \BibitemOpen
  \bibfield  {author} {\bibinfo {author} {\bibfnamefont {R.}~\bibnamefont
  {Barczyk}}, \bibinfo {author} {\bibfnamefont {S.}~\bibnamefont {Nechayev}},
  \bibinfo {author} {\bibfnamefont {M.~A.}\ \bibnamefont {Butt}}, \bibinfo
  {author} {\bibfnamefont {G.}~\bibnamefont {Leuchs}},\ and\ \bibinfo {author}
  {\bibfnamefont {P.}~\bibnamefont {Banzer}},\ }\href@noop {} {\bibfield
  {journal} {\bibinfo  {journal} {Phys. Rev. A}\ }\textbf {\bibinfo {volume}
  {99}},\ \bibinfo {pages} {063820} (\bibinfo {year} {2019})}\BibitemShut
  {NoStop}%
\bibitem [{\citenamefont {Debnath}\ and\ \citenamefont
  {Viswanathan}(2020{\natexlab{a}})}]{CLEO2020}%
  \BibitemOpen
  \bibfield  {author} {\bibinfo {author} {\bibfnamefont {A.}~\bibnamefont
  {Debnath}}\ and\ \bibinfo {author} {\bibfnamefont {N.~K.}\ \bibnamefont
  {Viswanathan}},\ }in\ \href
  {http://www.osapublishing.org/abstract.cfm?URI=CLEO_QELS-2020-JTh2E.1} {\emph
  {\bibinfo {booktitle} {Conference on Lasers and Electro-Optics}}}\ (\bibinfo
  {publisher} {Optical Society of America},\ \bibinfo {year} {2020})\ p.\
  \bibinfo {pages} {JTh2E.1}\BibitemShut {NoStop}%
\bibitem [{\citenamefont {Debnath}\ and\ \citenamefont
  {Viswanathan}(2021)}]{ADNKVBrew2021}%
  \BibitemOpen
  \bibfield  {author} {\bibinfo {author} {\bibfnamefont {A.}~\bibnamefont
  {Debnath}}\ and\ \bibinfo {author} {\bibfnamefont {N.~K.}\ \bibnamefont
  {Viswanathan}},\ }\href@noop {} {\bibfield  {journal} {\bibinfo  {journal}
  {Phys. Rev. A}\ }\textbf {\bibinfo {volume} {103}},\ \bibinfo {pages}
  {013510} (\bibinfo {year} {2021})}\BibitemShut {NoStop}%
\bibitem [{\citenamefont {Debnath}\ and\ \citenamefont
  {Viswanathan}(2020{\natexlab{b}})}]{ADNKVrt2020}%
  \BibitemOpen
  \bibfield  {author} {\bibinfo {author} {\bibfnamefont {A.}~\bibnamefont
  {Debnath}}\ and\ \bibinfo {author} {\bibfnamefont {N.~K.}\ \bibnamefont
  {Viswanathan}},\ }\href@noop {} {\bibfield  {journal} {\bibinfo  {journal}
  {J. Opt. Soc. Am. A}\ }\textbf {\bibinfo {volume} {37}},\ \bibinfo {pages}
  {1971} (\bibinfo {year} {2020}{\natexlab{b}})}\BibitemShut {NoStop}%
\bibitem [{\citenamefont {Pancharatnam}(1956)}]{P1956}%
  \BibitemOpen
  \bibfield  {author} {\bibinfo {author} {\bibfnamefont {S.}~\bibnamefont
  {Pancharatnam}},\ }\href@noop {} {\bibfield  {journal} {\bibinfo  {journal}
  {Proc. Ind. Acad. Sci. A}\ }\textbf {\bibinfo {volume} {44}},\ \bibinfo
  {pages} {247} (\bibinfo {year} {1956})}\BibitemShut {NoStop}%
\bibitem [{\citenamefont {Berry}(1984)}]{Berry1984}%
  \BibitemOpen
  \bibfield  {author} {\bibinfo {author} {\bibfnamefont {M.~V.}\ \bibnamefont
  {Berry}},\ }\href@noop {} {\bibfield  {journal} {\bibinfo  {journal} {Proc.
  R. Soc. A}\ }\textbf {\bibinfo {volume} {392}},\ \bibinfo {pages} {45}
  (\bibinfo {year} {1984})}\BibitemShut {NoStop}%
\bibitem [{\citenamefont {Berry}(1987)}]{Berry1987}%
  \BibitemOpen
  \bibfield  {author} {\bibinfo {author} {\bibfnamefont {M.~V.}\ \bibnamefont
  {Berry}},\ }\href@noop {} {\bibfield  {journal} {\bibinfo  {journal} {J. Mod.
  Opt.}\ }\textbf {\bibinfo {volume} {34}},\ \bibinfo {pages} {1401} (\bibinfo
  {year} {1987})}\BibitemShut {NoStop}%
\bibitem [{\citenamefont {Shapere}\ and\ \citenamefont
  {Wilczek}(1989)}]{Shapere}%
  \BibitemOpen
  \bibfield  {author} {\bibinfo {author} {\bibfnamefont {A.}~\bibnamefont
  {Shapere}}\ and\ \bibinfo {author} {\bibfnamefont {F.}~\bibnamefont
  {Wilczek}},\ }\href@noop {} {\emph {\bibinfo {title} {Geometric Phases in
  Physics}}}\ (\bibinfo  {publisher} {World Scientific, Singapore},\ \bibinfo
  {year} {1989})\BibitemShut {NoStop}%
\bibitem [{\citenamefont {Bliokh}\ \emph {et~al.}(2008)\citenamefont {Bliokh},
  \citenamefont {Gorodetski}, \citenamefont {Kleiner},\ and\ \citenamefont
  {Hasman}}]{Bliokh2008}%
  \BibitemOpen
  \bibfield  {author} {\bibinfo {author} {\bibfnamefont {K.~Y.}\ \bibnamefont
  {Bliokh}}, \bibinfo {author} {\bibfnamefont {Y.}~\bibnamefont {Gorodetski}},
  \bibinfo {author} {\bibfnamefont {V.}~\bibnamefont {Kleiner}},\ and\ \bibinfo
  {author} {\bibfnamefont {E.}~\bibnamefont {Hasman}},\ }\href@noop {}
  {\bibfield  {journal} {\bibinfo  {journal} {Phys. Rev. Lett.}\ }\textbf
  {\bibinfo {volume} {101}},\ \bibinfo {pages} {030404} (\bibinfo {year}
  {2008})}\BibitemShut {NoStop}%
\bibitem [{\citenamefont {Bliokh}(2009)}]{Bliokh2009}%
  \BibitemOpen
  \bibfield  {author} {\bibinfo {author} {\bibfnamefont {K.~Y.}\ \bibnamefont
  {Bliokh}},\ }\href@noop {} {\bibfield  {journal} {\bibinfo  {journal} {J.
  Opt. A: Pure Appl. Opt.}\ }\textbf {\bibinfo {volume} {11}},\ \bibinfo
  {pages} {094009} (\bibinfo {year} {2009})}\BibitemShut {NoStop}%
\bibitem [{\citenamefont {Bliokh}\ \emph {et~al.}(2010)\citenamefont {Bliokh},
  \citenamefont {Alonso}, \citenamefont {Ostrovskaya},\ and\ \citenamefont
  {Aiello}}]{BA2010}%
  \BibitemOpen
  \bibfield  {author} {\bibinfo {author} {\bibfnamefont {K.~Y.}\ \bibnamefont
  {Bliokh}}, \bibinfo {author} {\bibfnamefont {M.~A.}\ \bibnamefont {Alonso}},
  \bibinfo {author} {\bibfnamefont {E.~A.}\ \bibnamefont {Ostrovskaya}},\ and\
  \bibinfo {author} {\bibfnamefont {A.}~\bibnamefont {Aiello}},\ }\href@noop {}
  {\bibfield  {journal} {\bibinfo  {journal} {Phys. Rev. A}\ }\textbf {\bibinfo
  {volume} {82}},\ \bibinfo {pages} {063825} (\bibinfo {year}
  {2010})}\BibitemShut {NoStop}%
\bibitem [{\citenamefont {Allen}\ \emph {et~al.}(1992)\citenamefont {Allen},
  \citenamefont {Beijersbergen}, \citenamefont {Spreeuw},\ and\ \citenamefont
  {Woerdman}}]{AllenOAM1992}%
  \BibitemOpen
  \bibfield  {author} {\bibinfo {author} {\bibfnamefont {L.}~\bibnamefont
  {Allen}}, \bibinfo {author} {\bibfnamefont {M.~W.}\ \bibnamefont
  {Beijersbergen}}, \bibinfo {author} {\bibfnamefont {R.~J.~C.}\ \bibnamefont
  {Spreeuw}},\ and\ \bibinfo {author} {\bibfnamefont {J.~P.}\ \bibnamefont
  {Woerdman}},\ }\href@noop {} {\bibfield  {journal} {\bibinfo  {journal}
  {Phys. Rev. A}\ }\textbf {\bibinfo {volume} {45}},\ \bibinfo {pages} {8185}
  (\bibinfo {year} {1992})}\BibitemShut {NoStop}%
\bibitem [{\citenamefont {Berry}(1998)}]{Berry1998}%
  \BibitemOpen
  \bibfield  {author} {\bibinfo {author} {\bibfnamefont {M.~V.}\ \bibnamefont
  {Berry}},\ }in\ \href@noop {} {\emph {\bibinfo {booktitle} {Proc. SPIE}}},\
  Vol.\ \bibinfo {volume} {3487}\ (\bibinfo {year} {1998})\ pp.\ \bibinfo
  {pages} {6--11}\BibitemShut {NoStop}%
\bibitem [{\citenamefont {Barnett}(2002)}]{Barnett2022}%
  \BibitemOpen
  \bibfield  {author} {\bibinfo {author} {\bibfnamefont {S.~M.}\ \bibnamefont
  {Barnett}},\ }\href@noop {} {\bibfield  {journal} {\bibinfo  {journal} {J.
  Opt. B}\ }\textbf {\bibinfo {volume} {4}},\ \bibinfo {pages} {S7} (\bibinfo
  {year} {2002})}\BibitemShut {NoStop}%
\bibitem [{\citenamefont {Goldstein}(2011)}]{Goldstein}%
  \BibitemOpen
  \bibfield  {author} {\bibinfo {author} {\bibfnamefont {D.~H.}\ \bibnamefont
  {Goldstein}},\ }\href@noop {} {\emph {\bibinfo {title} {Polarized Light}}},\
  \bibinfo {edition} {3rd}\ ed.\ (\bibinfo  {publisher} {CRC Press, Taylor \&
  Francis Group, FL},\ \bibinfo {year} {2011})\BibitemShut {NoStop}%
\bibitem [{\citenamefont {Ghai}\ \emph {et~al.}(2009)\citenamefont {Ghai},
  \citenamefont {Senthilkumaran},\ and\ \citenamefont {Sirohi}}]{SingleSlit}%
  \BibitemOpen
  \bibfield  {author} {\bibinfo {author} {\bibfnamefont {D.~P.}\ \bibnamefont
  {Ghai}}, \bibinfo {author} {\bibfnamefont {P.}~\bibnamefont
  {Senthilkumaran}},\ and\ \bibinfo {author} {\bibfnamefont {R.~S.}\
  \bibnamefont {Sirohi}},\ }\href@noop {} {\bibfield  {journal} {\bibinfo
  {journal} {Opt Laser Eng}\ }\textbf {\bibinfo {volume} {47}},\ \bibinfo
  {pages} {123} (\bibinfo {year} {2009})}\BibitemShut {NoStop}%
\end{thebibliography}%


%

\end{document}